\documentclass[openacc]{rstransa}

\usepackage{mathrsfs}
\usepackage{bm}
\usepackage{soul}
\usepackage{mathtools}
\usepackage{laussy}
\newcommand{\av}[1]{\langle  #1 \rangle}
\newcommand{\mean}[1]{\langle#1\rangle}
\newcommand{\g}[1]{g^{(#1)}}

\titlehead{Research}

\begin{document}

\title{Two photons everywhere}

\author{
E. Zubizarreta Casalengua$^{1}$, F. P. Laussy$^{2}$ and E. del Valle$^{3}$}

\address{
$^{1}$Walter Schottky Institute, School of Computation, Information and Technology and MCQST, Technische Universit\"at M\"unchen, 85748 Garching, Germany\\
$^{2}$Instituto de Ciencia de Materiales de Madrid ICMM-CSIC, 28049 Madrid, Spain\\
$^{3}$Departamento de F\'isica Te\'orica de la Materia Condensada e IFIMAC, Universidad Aut\'onoma de Madrid, 28049 Madrid, Spain.\\Institute for Advanced Study, Technische Universit\"at M\"unchen, 85748 Garching, Germany.
}

\subject{Quantum Optics, Multiphoton sources}


\corres{Elena del Valle\\
\email{elena.delvallereboul@gmail.com}}

\begin{abstract}
  We discuss two-photon physics, taking for illustration the
  particular but topical case of resonance fluorescence. We show that
  the basic concepts of interferences and correlations provide at the
  two-photon level an independent and drastically different picture
  than at the one-photon level, with landscapes of correlations that
  reveal various processes by spanning over all the possible
  frequencies at which the system can emit. Such landscapes typically
  present lines of photon bunching and circles of antibunching. The
  theoretical edifice to account for these features rests on two
  pillars: i) a theory of frequency-resolved photon correlations and
  ii) admixing classical and quantum fields.  While experimental
  efforts have been to date concentrated on correlations between
  spectral peaks, strong correlations exist between photons emitted
  away from the peaks, which are accessible only through multiphoton
  observables. These could be exploited for both fundamental
  understanding of quantum-optical processes as well as applications
  by harnessing these unsuspected resources.
\end{abstract}

\begin{fmtext}
  \section{Introduction}
  Quantum mechanics is notorious for its quantized spectral
  lines. This is how the theory was born and the fact after which it
  is named, following Bohr~\cite{bohr13a}'s quantization of angular
  momentum of the electron orbitals. This produced Rydberg's constant
  as a product of fundamental constants and explained the lines as
  \phantom{somethingyouwanttohide}
\end{fmtext}

\maketitle

\begin{figure}
  \centering\includegraphics[width=.85\linewidth]{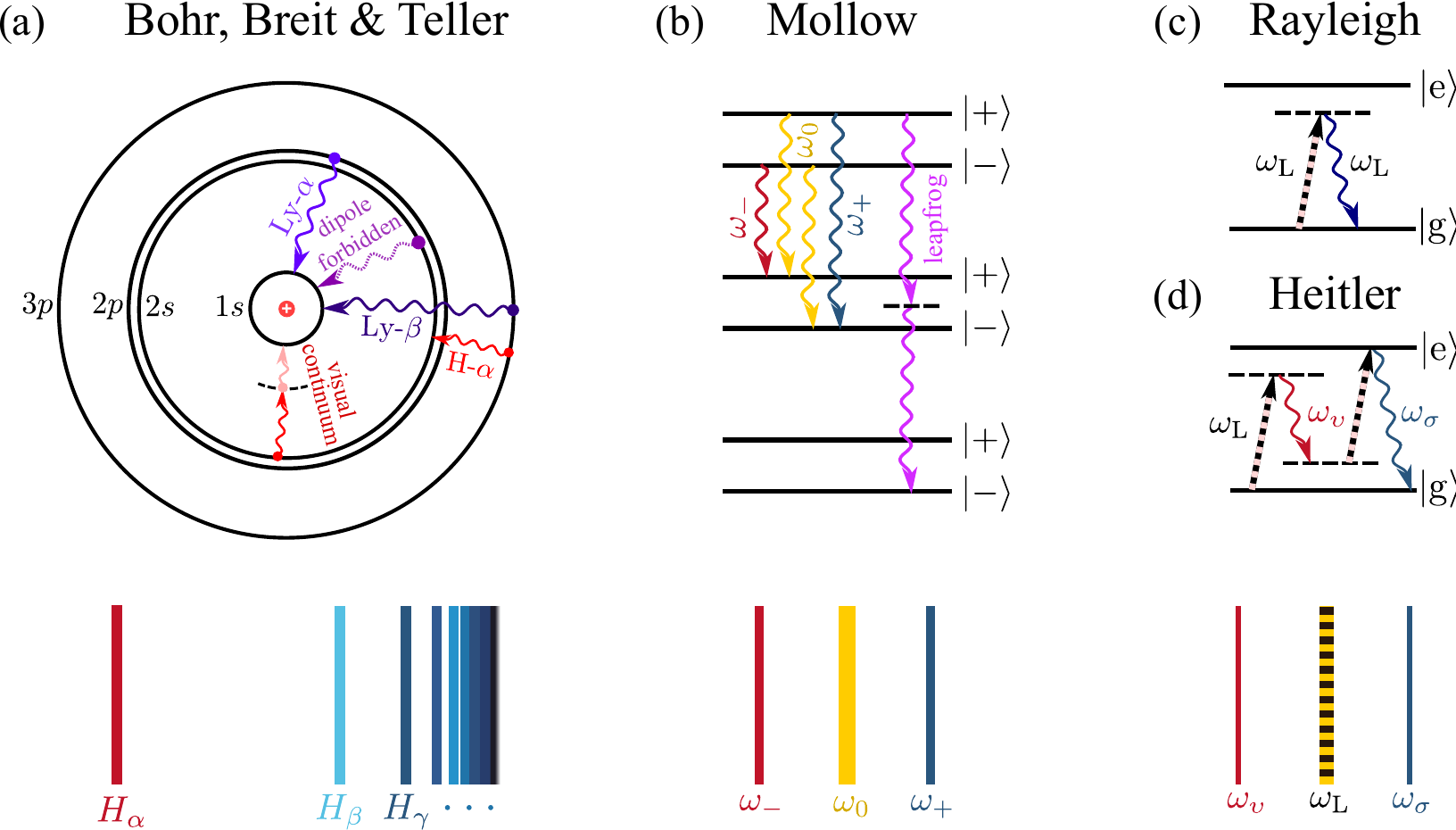}
  \caption{(a) Old quantum mechanics: transitions between quantized
    orbitals of the electron lead to a quantized spectrum. Transitions
    to~$n=2$ form the visible (Balmer) series. The $2\to1$ (Lyman) is
    in the ultraviolet. The $2s\to1s$ is furthermore dipole forbidden,
    but it can occur with a continuum of two-photon emission, which
    falls again in the visible range. (b) Transitions between dressed
    states~$\ket{\pm}$ lead to a triplet spectrum (Mollow). A
    two-photon, so-called \emph{leapfrog}, transition can also occur
    that jumps over an intermediate state. However weak is this
    transition, it can be revealed by two-photon correlations. (c) In
    the low-driving limit, detuning makes the same physics take
    another turn, with the triplet now formed by the Rayleigh
    scattering as the central peak and (d) breaking the leapfrog over
    one real-state transition but keeping the other virtual,
    accounting for the two side peaks. This results in a rich
    two-photon physics beyond the side peaks, not visible in
    photoluminescence but revealed in the two-photon correlation
    spectrum.}
  \label{fig:Fri9Feb094747CET2024}
\end{figure}

\noindent transitions between energy levels
(Fig.~\ref{fig:Fri9Feb094747CET2024}(a)).  At about the same time, a
much less resounding finding was recorded in a bulletin of the Lowell
Observatory: a nebula in the Pleiades was observed to have a
completely continuous spectrum of emission~\cite{slipher12a}.  There
are many ways to produce a continuous spectrum, from unbound charges
to, as was the case here, scattering from a broad spectrum (the light
from the star Merope).  Systematic and careful observations revealed
that emission nebulae (in particular planetary nebulae) which do not
reflect but emit light directly, also feature a continuous spectrum in
addition to the spectral lines of what was by then well-established
quantum theory~\cite{page36a}. One part---the Balmer continuum---could
be explained with the known mechanisms from reflection nebulae, but on
the other side of the Balmer limit, sitting with the quantized
spectral lines in the visible spectrum, lied a continuum spectrum
which was of unknown origin~\cite{zanstra36a}. This got simultaneously
and independently resolved in 1950, from both sides of the iron
curtain~\cite{kipper50a,spitzer51a}, as a fitting companion to Bohr’s
model of quantum jumps in the form of a two-photon emission from the
metastable $2S_{1/2}$ shell of the hydrogen atom down to the ground
state~$1S_{1/2}$ (Fig.~\ref{fig:Fri9Feb094747CET2024}(a)). Would the
atom be excited in the $2P$ level, it would undergo a normal,
``Bohrian'' transition back to the ground state, but the $2S$--$1S$ is
dipole forbidden so it has to go through the next (quadrupole) order,
which is however very weak, and, in ambient conditions, is knocked-off
by collision back into a radiative state. In the extremely rarefied
cosmic conditions, however, there is plenty of space as well as enough
time to have isolated Hydrogen atoms left stuck in great quantity in
their~$2S$ state. For those, the next best route back to~$1S$ is the
``Doppelemission'' (double-emission or two-photon emission) theorized
by Goeppert Mayer~\cite{goppertmayer29a} and first applied to Hydrogen
by Breit and Teller~\cite{breit40a}.  Although a higher-order process,
it is not dipole forbidden and finds in outer space the ideal quiet
laboratory conditions making possible its direct observation, indeed
giving it the name of ``visual continuum''. The Earth-based laboratory
is less auspicious for a direct (visual) observation of two-photon
emission~\cite{lipeles65a} but the process nevertheless opened the
multiphoton page of atomic
physics~\cite{zalialiutdinov18a}. 
Such two-photon transitions have, among other things, be generalized
to link any~$(n,l,m)$ states of the atom~\cite{tung84a} and can be
stimulated with a laser~\cite{yatsiv68a}.  In the solid state, the
opposite regime of the interstellar one can be realized of high
populations in small volumes, e.g., in semiconductors, where
delocalized electrons in the crystal under high laser excitation can
jump the bandgap in sufficient amount to produce a measurable and even
controllable two-photon continuous spectrum~\cite{hayat08a}. Another
solid-state approach is to inflate the light-matter coupling by
focusing light onto the emitter, to make higher-order processes ``less
smaller'' and thus lift their ``forbidden'' character, possibly making
them even comparable to first-order processes~\cite{rivera16a}.

Here, we present a quantum optical alternative to the QED description.
The latter typically relies on a perturbative treatment of the
two-photon processes of a complex system: the Hydrogen atom in its
simplest case, up to an interstellar ionized gas bathed in the
radiation of other astronomical objects. Instead, we consider the
\emph{exact} treatment of a \emph{simple} system, namely, resonance
fluorescence, i.e., the photon emission (fluorescence) of a two-level
system driven at the same energy as it is excited (resonance).  This
will allow us to focus on the two-photon physics itself, instead of
interesting but secondary problems specific to Hydrogen or to the
thermodynamics of nebulae. While we invoke a variety of our results
collected over the last decade, we try to keep the discussion
self-contained with no need of prior familiarity with our earlier
works on, mainly, frequency-resolved multiphoton
correlations~\cite{delvalle12a} or multiphoton interferences of
quantum fields~\cite{zubizarretacasalengua20a}.  We furthermore
connect these two aspects to provide a new and fairly comprehensive
picture of the phenomenology of two-photon emission from resonance
fluorescence, explaining features hitherto only observed. 
Section~\ref{sec:Sat10Feb185005CET2024} introduces the textbook
problem of resonance fluorescence but revisited with the ``sensor
formalism''~\cite{delvalle12a}, which is an alternative way to compute
spectra that will allow us, in Section~\ref{sec:Sun7Jan122947CET2024},
to provide a first departure from conventional treatments, by
introducing our concept of frequency-resolved correlations, which will
put the two-photon observables on the same footing as the one-photon
ones, and explain the reason for our title of ``two photons
everywhere''. In Section~\ref{sec:Sun7Jan155552CET2024}, we introduce
the other pillar of our edifice: interferences of quantum fields, as
underpinning their statistical
properties~\cite{zubizarretacasalengua20a}.  The basic idea, that
holds with quantum states, is applied to a dynamical system in
Section~\ref{sec:Sun7Jan163313CET2024}, which, in spirit, is resonance
fluorescence itself, but we show that the two-photon observables are
actually captured by the simpler case of a squeezed cavity, allowing
us to identify what is specific to quantum states admixtures of
Gaussian states (squeezing and coherent states) and what is to the
``more quantum'' two-level system, to which we return in
Section~\ref{sec:Mon19Feb221640CET2024} with a focus on detuning.  We
provide a surprisingly compact expression for two-photon correlations
in this case, that is exact to leading order in the driving and
captures its main phenomenology.  In
Section~\ref{sec:Sun11Feb211557CET2024}, we show how our approach
gives way to countless variations, even if remaining at the level of
resonance fluorescence, although this could and should be extended to
all possible quantum emitters. Specifically, we consider entanglement
and other quantum resources such as two-mode squeezing, which we
merely exhibit but that could be similarly explained and
exploited. Our approach could also usefully revisit closely related
problems such as two-photon absorption~\cite{loudon84a}, two-photon
resonance fluorescence~\cite{holm85a} or two-photon
gain~\cite{lewenstein90a}.

\section{Resonance Fluorescence}
\label{sec:Sat10Feb185005CET2024}

\begin{figure}[!h]
  \centering\includegraphics[width=.9\linewidth]{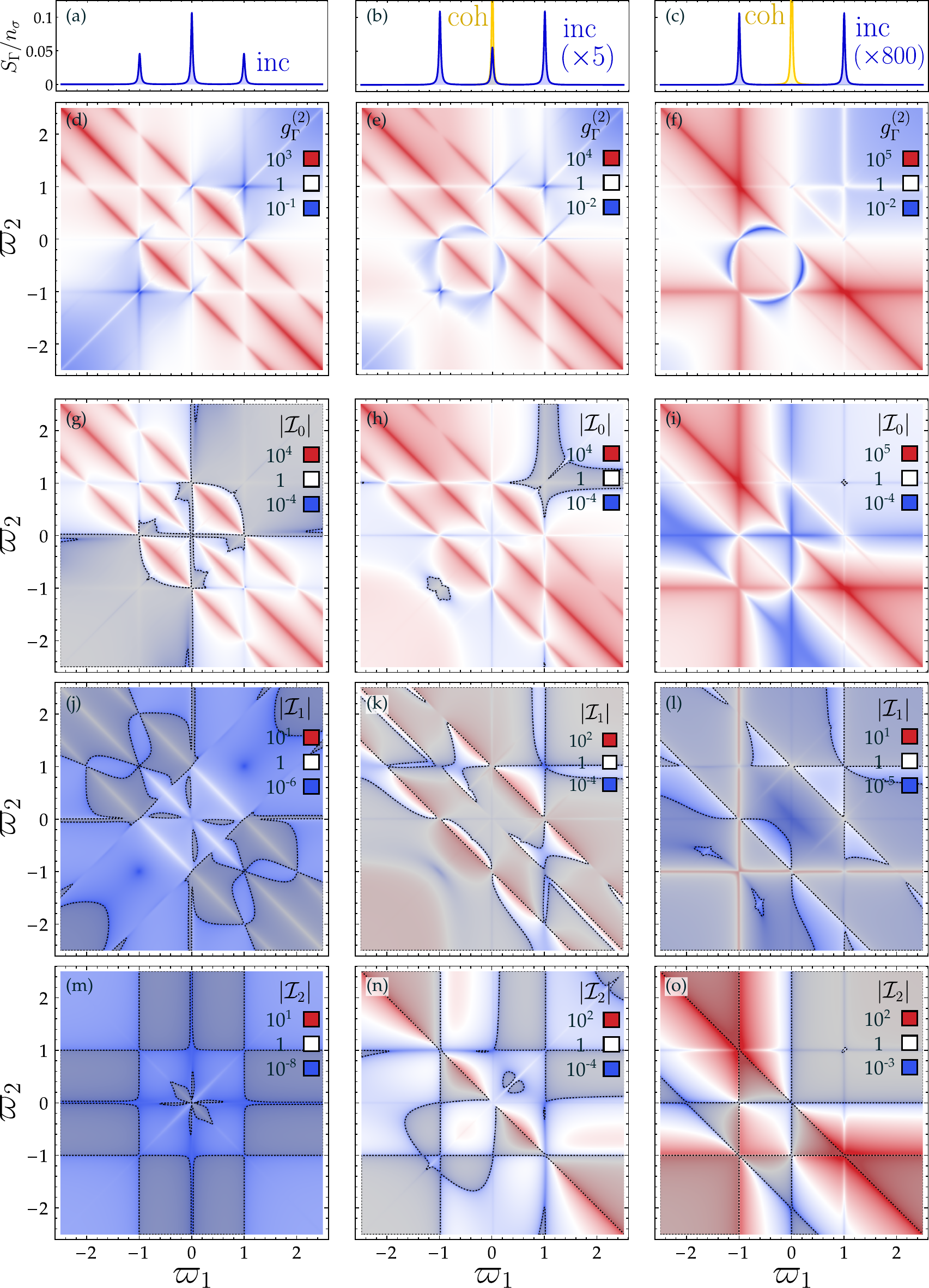}
  \caption{The two-photon physics of resonance
    fluorescence. \textbf{Top row:} photoluminescence spectra for (a)
    the Mollow triplet at resonance, (b) the detuned Mollow triplet
    and (c) the detuned Heitler triplet. The coherent (scattered)
    light is shown in yellow.  \textbf{Second row:} two-photon
    correlation spectra $g^{(2)}_\Gamma(\varpi_1,\varpi_2)$ for the
    corresponding spectra, with bunching in red, no-correlation in
    white and antibunching or no-coincidence emission in blue.  Two
    sets of features dominate the landscape: straight lines of
    bunching (red) and circles of antibunching (blue). Lines result
    from multiphoton transitions that involve virtual
    photons---``leapfrog processes''---while circles result from
    self-homodyning destructive interferences.  \textbf{Three bottom
      rows}: covariance~$\mathcal{I}_0$ (third row), anomalous
    two-photon moments~$\mathcal{I}_1$ (fourth) and
    squeezing~$\mathcal{I}_2$ (bottom) which sum together with~1 to
    the two-photon spectrum. Shaded regions refer to negative
    quantities. Parameters: $\gamma_\sigma=1$ (setting the unit) and
    $\Gamma=2$ everywhere, while $(\Delta_\sigma, \Omega_\sigma)=$ (a)
    $(0, 40.05)$, (b) $(60,26.53)$ and~(c) $(80,2)$.}
\label{fig:Sun7Jan183134CET2024}
\end{figure}

Resonance fluorescence is the simplest problem of quantum optics, yet
a still actively investigated one.  Its Hamiltonian in the rotating
frame of the laser is (with $\hbar=1$)
\begin{equation}
  \label{eq:Sun7Jan201422CET2024}
  H_\sigma\equiv \Delta_\sigma \, \ud{\sigma}\sigma + \Omega_\sigma (\ud{\sigma} + \sigma)
\end{equation}
where~$\sigma$ is the annihilation operator of a two-level system
driven coherently by a laser with
amplitude~$\Omega_\sigma\in\mathbb{R}$, and $\Delta_\sigma$ is the
detuning from the laser frequency~$\omega_\mathrm{L}$ which we shall
take as the reference, i.e. $\omega_\mathrm{L}=0$.  This problem
provides a rich quantum-optical playground, from essentially two
regimes of excitation: at low-driving (so-called Heitler
regime~\cite{heitler_book54a}), the spectrum is dominated by the
Rayleigh-scattered light of the laser, while at high-driving
(so-called Mollow regime~\cite{mollow69a}), there is a splitting of
the spectral shape into a triplet
(Fig.~\ref{fig:Fri9Feb094747CET2024}). While these two regimes are
very different in character, we can find a first unifying theme
through large detuning. When the laser drives the system far from its
resonance frequency (although still referring to ``resonance
fluorescence''), the spectral response remains a symmetric
triplet. This is shown at the top of
Fig.~\ref{fig:Sun7Jan183134CET2024} for (a) the Mollow triplet at
resonance, (b) the detuned Mollow triplet with an increasingly bright
coherent peak sitting on top of a dim fluorescent one and (c) the
detuned (Heitler regime of) resonance fluorescence with vanishing
fluorescent contributions. {The driving~$\Omega_\sigma$ has been
  chosen} so that the spectral position of the side peaks is the same
in units of the Mollow
splitting~$\Omega_+ \approx \sqrt{\Delta_\sigma^2 + 4
  \Omega_\sigma^2}$ (at large enough detunings). The
frequency~$\omega$ is normalized to this splitting, i.e.,
$\varpi\equiv(\omega-\omega_\mathrm{L})/\Omega_+$. The central peak is
either (a) a fluorescent (broad) peak, (c) the Rayleigh-scattered
coherent peak (intense and narrow central peak, note that the side
peaks are magnified by a factor~$800$) or (b) an admixture of a
fluorescent and coherent peaks.

One can conveniently derive such results with our sensor
formalism~\cite{delvalle12a}, which is an alternative to formal
results that rely on mathematical structures of the problem---such as
the quantum regression theorem or the Wiener-Khinchin theorem---to
obtain instead observables from a physical modeling of what is being
measured. This was developed to extend the physical spectrum of Eberly
and W\'odkiewicz~\cite{eberly77a} to~$N$-photon observables, which
instead of computing awkward multi-time, normally-ordered integrals,
simply attaches to the system ``sensors'' (two of them for two-photon
correlations, $N$ sensors in general) and computes correlations
directly from their usual quantum averages.  Technically, this
requires to ``plug'' to the Hamiltonian of the system (in our case,
Eq.~(\ref{eq:Sun7Jan201422CET2024})), the sensors, themselves most
simply described as two-level systems~$\varsigma_i$ whose
frequencies~$\omega_i$ define which frequencies are being
``measured''. The name ``sensor'' was chosen as opposed to
``detector'' since correlations are obtained in the
limit~$\epsilon\to0$ of their vanishing coupling to the system,
thereby not affecting its dynamics. This restricts their use to
correlations as opposed to signal. Alternatively, one can also use the
``cascading systems''~\cite{gardiner93a,carmichael93b}, which was
shown in Ref.~\cite{lopezcarreno18a} to be equivalent to the sensor
method and to conveniently substitute it in cases where the signal is
needed, e.g., to perform frequency-resolved Monte Carlo simulations.
For resonance fluorescence, the augmented Hamiltonian thus reads:
\begin{equation}
  \label{eq:Sun7Jan163806CET2024}
  H_{\sigma;\varsigma} \equiv H_\sigma + \Delta_1 \ud{\varsigma_1}\varsigma_1 + \Delta_2 \ud{\varsigma_2} \varsigma_2
  + \epsilon \sum_{i = 1,2} (\ud{\sigma} \varsigma_i + \ud{\varsigma_i} \sigma )\,.
\end{equation}
In the rotating frame of the laser, the
frequencies~$\Delta_i\equiv\omega_i-\omega_\mathrm{L}$ are
simply~$\omega_i$ since~$\omega_\mathrm{L}=0$.  As a quantum-optical
problem, one should include dissipation, which can be provided by a
master equation in the Lindblad form, so that the dynamics for the
full density matrix~$\rho$---of the system itself as well as its
sensors---is governed by the equation
\begin{equation}
  \label{eq:Sun7Jan203904CET2024}
  \partial_t{\rho} = -i [H_{\sigma},\rho] + {\gamma_\sigma\over2}\mathcal{L}_\sigma\rho+ \sum_{j = 1, 2}
  \frac{\gamma_j}{2} \mathcal{L}_{\varsigma_j} \rho,
\end{equation}
where the Lindblad terms are of the form
$\mathcal{L}_c\rho \equiv 2 c \rho \ud{c} - \ud{c} c \rho - \rho
\ud{c} c$ for any operator~$c$. Importantly, in addition to the decay
rate~$\gamma_\sigma$ of the two-level system, we also bring the decay
rate~$\gamma_i$ of the $i$th sensor, that describes its frequency
bandwidth.  For two-photon observables (and higher~$N$), such a
parameter is mandatory for a physical description of the system,
unlike one-photon observables like the power spectrum, which can be
well described for a vanishing linewidth of their detector (recovering
the Wiener-Khinchin result).  In the following we shall assume the
same frequency window for both sensors, i.e., $\gamma_{1,2} = \Gamma$,
which we shall refer to as the ``filter width'' as this also
corresponds to placing an interference filter before an ideal
detector.  The sensor method allows us to obtain the luminescence
spectrum
$S_\Gamma(\omega)={\Gamma\over2\pi\epsilon^2}\langle\ud{\varsigma}\varsigma\rangle(\omega)$
(no index needed since a single sensor is enough) as the population of
the sensor~$\varsigma$ placed at the frequency~$\omega$, which is a
mere parameter in Eq.~(\ref{eq:Sun7Jan163806CET2024}) as opposed to a
physical observable and thus an operator, which would complicate very
much the treatment. In particular, note that we need not invoke any
quantum-regression theorem, Fourier transform, etc. For resonance
fluorescence, the spectral shape has a simple, characteristic
structure:
\begin{equation}
  \label{eq:Wed14Feb000029CET2024}
  S_\Gamma(\omega)=|\av{\sigma}|^2\mathscr{L}_\Gamma(\omega)+\big(\langle\ud{\sigma}\sigma\rangle-|\av{\sigma}|^2\big)\mathscr{M}_\Gamma(\omega)
\end{equation}
where
\begin{equation}
  \label{eq:Wed14Feb000415CET2024}
  \mean{\sigma} = \frac{2 \Omega_\sigma (2 \Delta_\sigma + i \, \gamma_\sigma)}{\gamma_\sigma^2 + 4 \Delta_\sigma^2 + 8 \Omega_\sigma^2} 
  \quad \text{and} \quad 
  \mean{\ud{\sigma} \sigma} = \frac{4 \Omega_\sigma^2}{\gamma_\sigma^2 + 4 \Delta_\sigma^2 + 8 \Omega_\sigma^2}
\end{equation}
are the coherently-scattered field~$\langle\sigma\rangle$ (with
intensity the modulus square of this) and the total
population~$\langle\ud{\sigma}\sigma\rangle$, including the coherent
part, so the incoherent part alone is
$\langle\ud{\sigma}\sigma\rangle-|\av{\sigma}|^2=32\Omega_\sigma^2/(\gamma_\sigma^2+4\Delta_\sigma^2+8\Omega_\sigma^2)^2$
(this happens to be $2\langle\ud{\sigma}\sigma\rangle^2$). These are
not $\Gamma$ dependent. They weight, respectively, a central
Lorentzian
peak~$\mathscr{L}_\Gamma(\omega)\equiv \frac{1}{\pi}\frac{\Gamma/2
}{(\Gamma/2)^2 + \omega^2}$ and a symmetric triplet (both normalized),
whose expression is a bit more involved:
\begin{multline}
  \label{eq:Wed14Feb001144CET2024}
  \mathscr{M}_\Gamma(\omega)\equiv\\
  {2\over\pi}{(\gamma_{12}^2+4\omega^2)(\gamma_{11}^2\gamma_{12}+4\gamma_{12}\Delta_\sigma^2+4\gamma_{10}\omega^2)+8\Omega_\sigma^2(\gamma_{11}\gamma_{12}\gamma_{35}+4\gamma_{12}\Delta_\sigma^2-4\gamma_{1\bar2}\omega^2)+128\Omega_\sigma^4\gamma_{11} \over 
\splitfrac{(\gamma_{12}^2+4\omega^2)\big[(\gamma_{11}^2+4\Delta_\sigma^2)^2+8(\gamma_{11}^2-4\Delta_\sigma^2)\omega^2+16\omega^4\big]+{}}{{}+32\Omega_\sigma^2\big[\gamma_{11}\gamma_{12}(\gamma_{11}^2+4\Delta_\sigma^2)+4(\gamma_{01}\gamma_{11}+4\Delta_\sigma^2)\omega^2-16\omega^4\big]+256\Omega_\sigma^4(\gamma_{11}^2+4\omega^2)}
}
\end{multline}
where we use the notation~$\gamma_{ij}\equiv i\Gamma+j\gamma_\sigma$
for any integer~$i$, $j$, with also~$\bar\jmath=-j$ so, e.g.,
$\gamma_{35}=3\Gamma+5\gamma_\sigma$. The expression is not
particularly enlightening but it is completely general, including the
Mollow triplet and Heitler regime, at and out-of resonance, also with
the effect of detection~$\Gamma$. An even more complete (with
incoherent pumping and dephasing) version of this expression for the
``physical spectrum'' of resonance fluorescence is available as a
self-standing applet~\cite{delvalle13b}. At any rate, its shape is
simple: it is a triplet which maintains a perfect symmetry around the
laser frequency set here at~$\omega_\mathrm{L}=0$, whose
interpretation in the various regimes of driving is given in
Figs.~\ref{fig:Fri9Feb094747CET2024}(b) and~(c). In both cases, Bohr's
insight provides a discerning picture of the otherwise mysterious
spectral features of resonance fluorescence.  At high-driving, the
so-called ``dressed atom'' picture considers such quantum jumps between
the dressed states
$\ket{\pm}\equiv c_\pm\ket{\mathrm{g}}\pm c_\mp\ket{\mathrm{e}}$ of a
two-level system~$\ket{\mathrm{g}}$ and~$\ket{\mathrm{e}}$ dressed by
photons from a driving laser,
with~$c_{\pm}\equiv(1+\xi^{\mp2})^{-1/2}$
and~$\xi\equiv\Omega_\sigma/[\sqrt{\Omega_\sigma^2+(\Delta_\sigma/2)^2}+(\Delta_\sigma/2)]$. The
central peak, at frequency~$\omega_0$, is twice as bright than the
side peaks $\omega_\pm$ due to two degenerate transitions
$\ket{+}\to\ket{+}$ and $\ket{-}\to\ket{-}$. At low driving, now
looking at the system in its bare states, a detuned laser does not
make it to the excited state and so is restrained chiefly to Rayleigh
(energy-conserving) scattering. However, involving two laser photons,
one can match the energy~$\omega_\sigma$ of the emitter with the
excess energy $\omega_v\equiv 2\omega_\mathrm{L}-\omega_\sigma$
(either smaller or larger than~$\omega_\sigma$ depending on detuning),
forming an exactly symmetric peak.  This perceptive physics is
well-known since the
80s~\cite{cohentannoudji77a,reynaud88a,dalibard83a}.  Its
understanding has little evolved since then~\cite{cohentannoudji16a}.

Since the photoluminescence spectrum is a single-photon observable,
one could consider a vanishing detector bandwidth~$\Gamma\to0$
(replacing~$\gamma_{ij}$ by $j\gamma_\sigma$), in which case one sees
a tightening of the lines, in particular, the coherent
Lorentzian~$\lim_{\Gamma\to0}\mathscr{L}_\Gamma(\omega)$ becomes a
Dirac~$\delta(\omega)$ function that reflects the vanishing linewidth
of the laser treated as a~$c$ number ($\Omega_\sigma$) in
Eq.~(\ref{eq:Sun7Jan201422CET2024}).  The effect of detection is thus
a mere (and expected) broadening of the lines, which does not
qualitatively alter the spectral shape.  The case of Heitler resonance
($\Omega_\sigma\ll\gamma_\sigma$ and~$\Delta_\sigma=0$) has attracted
deserved but unwary attention. A short discussion will allow us to
motivate the significance of detection, to which the sensor formalism
gives utmost importance.  The
spectrum~(\ref{eq:Wed14Feb001144CET2024}) reduces in this case~to:
\begin{equation}
  \label{eq:Thu15Feb115918CET2024}
  S_\Gamma(\omega)={4\Omega_\sigma^2\over\gamma_\sigma^2}\left[\left(1-{16\Omega_\sigma^2\over\gamma_\sigma^2}\right)\mathscr{L}_\Gamma(\omega)+\frac{8 \Omega_\sigma^2}{\gamma_\sigma^2} \frac{2}{\pi} \frac{\gamma_{12} \gamma_{11}^2 + 4 \Gamma \omega^2}{(\gamma_{11}^2 + 4 \omega^2)^2}\right] 
\end{equation}
and, again, the $\Gamma\to0$ appears to be a straightforward and
good-enough approximation of the main features
\begin{equation}
  \label{eq:Thu15Feb092755CET2024}
   S_0(\omega)={4\Omega_\sigma^2\over\gamma_\sigma^2}\left[\left(1-{16\Omega_\sigma^2\over\gamma_\sigma^2}\right)\delta(\omega)+{8\Omega_\sigma^2\over\gamma_\sigma^2}{\pi\gamma_\sigma}\mathscr{L}_{\gamma_\sigma}^2(\omega)\right] 
\end{equation}
in fact making them particularly transparent, namely, as long
as~$\Gamma\le\gamma_\sigma$ (sub-natural linewidth resolution), given
that~$\Omega_\sigma^2\ll\gamma_\sigma^2$, the spectrum is essentially
a very narrow central line sitting on a weak and broad
($\gamma_\sigma$) spectrum. This scenario can be seen for the central
peak in panel~(b) of Fig.~\ref{fig:Sun7Jan183134CET2024}. The
fluorescent spectrum is, to leading order, the \emph{square} of a
Lorentzian, exposing its two-photon origin. This expression corrects
the erroneous Eq.~(3) in Ref.~\cite{lopezcarreno18b}. The impact of
the detector in Eq.~(\ref{eq:Thu15Feb092755CET2024}) is merely to
contribute a quite innocuous broadening: one can take better and
better detectors and converge to the ideal
limit~(\ref{eq:Thu15Feb092755CET2024}) which is essentially
${4\Omega_\sigma^2\over\gamma_\sigma^2}\delta(\omega)$, so it would
appear that we are merely discussing technical issues. That is so, at
the one-photon level.

Two-photon observables, however, behave differently, at least, if
taken in their entirety. The complete quantity should retain both the
frequencies~$\omega_1$ and~$\omega_2$ as well as the times of
detection~$t_1$ and~$t_2$ of both photons, or, in a steady state, the
time difference~$\tau\equiv t_2-t_1$. We will introduce it in the next
Section, but for now, to compare with the results found in the
literature, we assume both frequencies to be that of the
emitter~$\omega_1=\omega_2=\omega_\sigma$, itself at resonance with
the laser frequency~$\omega_\mathrm{L}$, in which case the sensor
formalism provides us with
$g^{(2}_\Gamma(\tau)=\big\langle\ud{\varsigma_1}(0)(\ud{\varsigma_2}\varsigma_2)(\tau)\varsigma_1(0)\big\rangle/[\langle\ud{\varsigma_1}\varsigma_1\rangle\langle\ud{\varsigma_2}\varsigma_2\rangle]$
which evaluates to
\begin{equation}
  \label{eq:Sun11Feb103038CET2024}
  \g{2}_{\Gamma} (\tau) = \left(e^{-(\Gamma+\gamma_\sigma)\tau/2}+{\Gamma\gamma_\sigma\over\Gamma^2-\gamma_\sigma^2}e^{-\Gamma\tau/2}-{\Gamma^2\over\Gamma^2-\gamma_\sigma^2}e^{-\gamma_\sigma\tau/2}\right)^2\,,
\end{equation}
(the limit~$\Gamma\to\gamma_\sigma$ gives
$\g{2}_{\gamma_\sigma}
(\tau)=\big[{1+(\gamma_\sigma\tau)/2\over2}e^{-\gamma_\sigma\tau/2}-e^{-\gamma_\sigma\tau}\big]^2$). The
most natural way to neglect frequencies is to detect them all, which
corresponds to taking the limit~$\Gamma\to\infty$, in which case one
recovers the result from the literature~\cite{carmichael76a}:
\begin{equation}
  \label{eq:Wed7Feb110037CET2024}
  \g{2}_{\infty} (\tau) = (1-e^{-\gamma_\sigma\tau/2})^2 \,.
\end{equation}
This describes an excellent single-photon source (in particular with a
flattening $\approx(\gamma_\sigma\tau/2)^2$ at small delays~$\tau$
around $\g{2}(0)=0$, characteristic of superior single-photon
emission~\cite{arXiv_zubizarretacasalengua23c}).  That is to say, one
apparently has a bright very
narrow---Eq.~(\ref{eq:Thu15Feb092755CET2024})---and
antibunched---Eq.~(\ref{eq:Wed7Feb110037CET2024})---source. This is
thanks to various ingredients contributing their respective benefits:
the laser brings the narrow linewidth while the two-level system
brings the single-photon emission. But in this listing of great
properties, which have been experimentally
demonstrated~\cite{nguyen11a,matthiesen12a}, one is using different
and in fact incompatible configurations, namely, $\Gamma\to0$ for the
spectrum and~$\Gamma\to\infty$ for the antibunching. In the
laboratory, separate characterizations, overlooking the impact
of~$\Gamma$, have been performed of the system, thereby omitting a
crucial adversative conjunction in describing an ``ultra-coherent
\emph{or} single photon source'' and ``subnatural linewidth \emph{or}
single photons''.  This oversight shows that~$\Gamma$ is not a mere
technical concern but a central consideration. If working directly
with Eqs.~(\ref{eq:Thu15Feb115918CET2024}) and
(\ref{eq:Sun11Feb103038CET2024}) instead of the textbook limits, then
the intrinsic limitations imposed by detection would be obvious.  The
difficulty is that one needs a large enough bandwidth~$\Gamma$ so as
to have sufficient integration time to resolve~$\g{2}(\tau)$ while
also having a small enough bandwidth to have a good frequency
resolution and not suffer too much broadening. When using the wrong
limit for each observable, one indeed gets the worst result possible,
namely, $S_\infty(\omega)$ vanishes as the spectrum becomes constant
over~$(-\infty,+\infty)$ since photons are emitted at all possible
energies, while $\g{2}_0(\tau)=1$ and all correlations are lost. These
are in fact those of an ideal laser; a physical laser has a linewidth
and filtering below that linewidth would go to the thermal limit
$\g{2}_0(\tau)=2!$~\cite{centenoneelen92a}, at any rate, the
single-photon correlations are completely lost. Presented in these
terms, it would seem that the time--frequency incompatibility occurs
at the Fourier level and is thus unavoidable. But this could be easily
circumvented merely by replacing the source with one that is more
spectrally narrow, i.e., by reducing~$\gamma_\sigma$, keeping the same
antibunching with a broader-than-natural linewidth of the substitute
emitter, but still narrower than the original one: in this way, one
can realize arbitrarily narrow and antibunched source, showing that
there is no fundamental limitation, only a technical one of finding
the suitable emitter.  In fact, even the sub-natural linewidth can be
realized \emph{simultaneously} with antibunching, provided a small,
but crucial, variation of the experiments~\cite{lopezcarreno18b}. One
first need to understand the nature of antibunching in this system,
which, as we said, comes from the two-level system. Or does it?
Because driving is weak in this regime, the saturation of the
two-level system is not actually needed, in contrast to incoherent
excitation where truncation of the number of excitations is crucial
(maintaining the average population below~$1/2$). Here, a weak
nonlinearity would work as well since the antibunching arises in this
case from a two-photon approximation of squeezing
antibunching~\cite{lemonde14a} (this is an approximation because there
is also three and higher-order antibunching which a squeezed state
cannot provide). Squeezing and antibunching have a long shared history
of joint appearances, initially with little appreciation of their
interconnectedness~\cite{mandel82a,arnoldus83b,loudon84a}. This is in
this way that they have been independently
theorized~\cite{stoler70a,dewitt_book64a},
sought~\cite{walls81a,carmichael76a} and ultimately
discovered~\cite{kimble77a,lu98a} in resonance fluorescence. In such a
system, they turn out to be two faces of the same coin, namely, of
wave interferences~\cite{carmichael85a}.  Interferences are at the
heart of both optics and quantum mechanics, so their importance in
quantum optics can only be understood as
momentous~\cite{ou_book07a,ficek_book04a}. While in classical optics,
the simplest waves are described by two parameters---their amplitude
and phase---quantum fields come with an infinite number of amplitudes
for non-Gaussian states of light, as each multiphoton
component~$\ket{n}$ of the field~$\ket{\psi}=\sum_{n}c_n\ket{n}$ comes
with its own independent coefficient~$c_n$. One can produce rich
multiphoton correlated outputs from admixing (or interfering) even the
simplest and most popular Gaussian
states~\cite{zubizarretacasalengua20b}, namely, the coherent
state~$\ket{\alpha}=\mathscr{D}(\alpha)\ket{0}$, defined in terms of
the displacement
operator~$\mathscr{D}(\alpha) \equiv \exp \big(\alpha \ud{a} -
\alpha^* a \big)$ where $\alpha = |\alpha| e^{i \phi}$ and~$a$ the
harmonic oscillator annihilation operator, so that
$\left(\mathscr{D}\right)^\dagger a \mathscr{D}= a +\alpha$, and the
single-mode (quadrature) squeezed state, defined in term of the
squeezing operator~\cite{loudon_book00a,carmichael_book02a}
$\mathcal{S}_1 \equiv \exp[\frac{1}{2} ( \xi^* a^2 - \xi \ud{a}^{2} )
]$.  The idea is as simple as for classical optics interferences: by
fine-tuning their respective coefficients, one can realize destructive
or constructive interferences for a given component~$\ket{n}$ and thus
suppress or maximize it in the output. At the Gaussian level, this
comes with strong constrains since few parameters define all the
multiphoton weights, but because such states are easily produced, they
have elicited most of the interest.  The idea of admixing squeezing
with a coherent state appeared very early, precisely with the aim to
produce antibunching~\cite{stoler74a} (under the name of
``anticorrelation effect''), and in fact even before the more
straightforward idea of a two-level system being restored into its
ground state~\cite{kimble76a}. It was then believed that such squeezed
anticorrelations might exist in the transient dynamics only. While
antibunching from the two-level system was quickly
observed~\cite{kimble77a}, the one based on squeezing took more time
(and with pulsed excitation) with the signal and pump of a degenerate
parametric amplifier, producing both bunching and antibunching by
varying the relative phase~\cite{koashi93a}.  The effect was more
clearly understood in terms of two-photon interferences by Ou \emph{et
  al.} who also realized it in the stationary regime~\cite{lu01a}.
Now that we understand better the origin of antibunching in resonance
fluorescence~\cite{hanschke20a} (as squeezing antibunching), we can
return to our hope of realizing it simultaneously with a subnatural
linewidth. The loss of antibunching~\cite{lopezcarreno22a} in
resonance fluorescence is due to the interference between the
incoherent and coherent components being disrupted by the detection,
which filters out the spectrally broad incoherent (squeezed) fraction
more than it does the coherent (narrow) one. One could thus restore
antibunching by correcting for this imbalance, which is an excess of
coherent state, therefore, destructive interferences with an external
laser can achieve that, as was indeed shown theoretically in
Ref.~\cite{lopezcarreno18b}. Such techniques that go by the name of
``homodyning'' have been demonstrated experimentally to extract the
quantum part of a signal~\cite{fischer16a}. Theoretically, this simply
consists in adding a complex field~$\alpha$ to the homodyned operator,
so $\sigma$ in our case, i.e., to substitute $\sigma\to\alpha+\sigma$
for a tunable~$\alpha\in\mathbb{C}$ in the Hamiltonian~$H_\sigma$. We
will go into the details of this mechanism in
Section~\ref{sec:Sun7Jan155552CET2024} where we will generalize it to
all frequencies of the system, while our current discussion is merely
its particular case for photons with the same frequency.  For now, it
will be enough to note that evidences of such interferences and their
effect on the correlations were reported in independent and
complementary works that tamper with either the incoherent fraction
(and thus loosing antibunching)~\cite{phillips20a,hanschke20a} or on
the opposite with the coherent fraction (producing excess bunching
instead)~\cite{masters23a}. Restoring antibunching remains to be
demonstrated experimentally.  At the theoretical level, our discussion
so far should have established that the detection, which manifests
itself through the parameter~$\Gamma$, is at the center of
quantum-optical characterizations and should not be treated
lightheartedly.  A deeper concept is concealed in
Eq.~(\ref{eq:Sun11Feb103038CET2024}), namely, that what is being
measured is truly the correlations of photons with
frequency~$\omega_\mathrm{L}=0$, i.e., one is making a \emph{complete}
characterization of both the time \emph{and} frequency of the photons,
which is why~$\Gamma$ is mandatory. What the textbook
limit~$\Gamma\to\infty$ really does is to neglect the frequency
information. This is not entirely apparent from
Eq.~(\ref{eq:Sun11Feb103038CET2024}) because it seems that no other
frequency than the one at which the system emits would make
sense. This is not the case.  Such a complete description for more
general multiphoton observables is discussed in the next Section,
which embarks us on another departure with the bulk of the literature,
this time not for a computational technique only, but at the
conceptual level of what it means to detect multiphotons.

\section{Frequency-resolved photon correlations}
\label{sec:Sun7Jan122947CET2024}

When considering the emission from a quantum emitter, and performing,
say, two-photon correlations from its output, there is an irresistible
temptation to correlate photons that are ``visible'' in the emission
spectrum, i.e., that originate from spectral peaks. If the emission
further comes in the form of several peaks, that invites for
cross-correlating them. This is due to the persistence of the
classical picture even to the quantum opticians, despite now many
decades of quantum theory telling us that quantum states at the
multi-particle level are not conditioned by their attributes at the
one-particle level~\cite{glauber63a}.  This makes it conceivable that
multi-photon correlations can be more pronounced or interesting in
spectral regions where the intensity (or population, i.e., a
one-photon observable of the type~$\av{\ud{a}a}$) is itself small
or even negligible. One must, at the quantum level, separate in
principle quantity and quality. It is in fact, beyond conceivable,
compulsory to elevate one's understanding of multiphoton emission to
such situations where the system does not emit at the one-photon
level, but does at the two-photon one, and vice-versa. There can also
be joint intense emission of the two type, or jointly suppressed. All
combinations are possible. They are therefore conceptually
disconnected. To see this, one needs a quantity to visualize
two-photon physics, which we now introduce. Such a quantity should
have no such prejudice for the spectral peaks, and treat every pair of
photons the same. It should be represented on a 2D plot as it requires
two axes, one for each photon. It should also extend or generalize
already existing quantities as it is unlikely that the essence of
two-photon physics has been missed entirely, although it might well
have been left incomplete. Such a two-photon quantity we call the
\emph{two-photon spectrum}~\cite{delvalle13a,gonzaleztudela13a} or
two-photon correlation spectrum, which we define in terms of the
intensity or population operator~$\hat n(\omega,t)$ that quantifies
the amount of quantum radiation at frequency~$\omega$ at
time~$t$~\cite{delvalle12a}, leading to:
\begin{equation}
  \label{eq:Thu15Feb190014CET2024}
  g^{(2)}_\Gamma(\omega_1,\omega_2,t_1,t_2)={\langle{:}\hat n(\omega_1,t_1)\hat n(\omega_2,t_2){:}\rangle\over\langle\hat n(\omega_1,t_1)\rangle\langle\hat n(\omega_2,t_2)\rangle}\,.
\end{equation}
This is a clear generalization of Glauber's two-photon correlation
function, but retaining the frequency information along with the
temporal one. For this reason, the detector's bandwidth~$\Gamma$ is
mandatory. This quantity has been considered by various
Authors~\cite{arnoldus84a,knoll86a,cresser87a} but it has presented
considerable difficulties and only partial results could be derived,
sometimes with inconsistencies.  From the sensor formalism, however,
it is easily and exactly obtained as
\begin{equation}
  \label{eq:Sun7Jan173131CET2024}
  g^{(2)}_{\Gamma}\left(\omega_1, \omega_2, t_1, t_2\right) = \frac{\big\langle{\ud{\varsigma_1}(t_1)(\ud{\varsigma_2} \varsigma_2)(t_2)\varsigma_1(t_1)}\big\rangle}{\big\langle{(\ud{\varsigma_1} \varsigma_1)(t_1)\big\rangle}{\big\langle(\ud{\varsigma_2} \varsigma_2)(t_2)}\big\rangle}\,.
\end{equation}
While one can consider both time and frequency (see, e.g.,
Refs.~\cite{gonzaleztudela15a,lopezcarreno17a}), it will be enough for
the present text to consider coincidences of a stationary state, i.e.,
the case~$t_1=t_2\to\infty$. This will not diminish the main insight
that in so doing, we are making a joint characterization in time
(albeit coincidences only) and frequencies.  We now embrace the full
picture and compute numerically two-frequency
coincidences~(\ref{eq:Sun7Jan173131CET2024}) through standard quantum
averages of the sensor method. Since frequencies are mere parameters,
this allows a convenient and exact computation of what was previously
obtained at the cost of great efforts and several oversimplifying
approximations~\cite{schrama92a,nienhuis93a}.  The results are shown
in Fig.~\ref{fig:Sun7Jan183134CET2024}'s panels~(d--f) [2nd row],
below the corresponding spectral shapes, reproducing three notable
cases from the pioneering work~\cite{gonzaleztudela13a}. Unlike the
top row, where~$S_\Gamma(\omega)$ could be obtained in the
limit~$\Gamma\to0$, these two-photon spectra would become ``trivial''
in this limit, with~$g^{(2)}_0(\omega_1,\omega_2)=1$, i.e., washing
out all traces of correlations as the detectors integrate over
infinite times. The other limit is more interesting but brings nothing
new as it recovers Eq.~(\ref{eq:Wed7Feb110037CET2024}), and so,
at~$\tau=0$, is identically zero. What is of interest, clearly, is the
intermediate case where a landscape of correlations is revealed.  It
will be enough to limit ourselves with a description of the
qualitative features, although we repeat that the results are
numerically exact, so one could study cross sections in more
quantitative details.  In our qualitative description, red colors
correspond to bunching, i.e., to photons with the respective
frequencies arriving together, with antidiagonals of bunching in
cases~(a) and~(b) that correspond to the two-photon Mollow
triplet. These lines indeed generalize Bohr's quantization condition
for transitions between two energies~$E_i-E_j=\omega$, to a two-photon
jump
\begin{equation}
  \label{eq:Sun7Jan225757CET2024}
  E_k-E_l=\omega_1+\omega_2
\end{equation}
where the energies~$E_k$ and~$E_l$ are non-contiguous in the energy
ladder and so the system effectively ``jumps over'' a real state, for
which reason this has been termed a \emph{leapfrog
  process}~\cite{delvalle13a,gonzaleztudela13a}. This is the direct
quantum-optical counterpart of G\"opert Mayer's Doppelemission with a
few but important variations. In her case, there was nothing to jump
over, and the multiphoton emission was thus in a different frequency
range, indeed, in its manifestation from nebulae emission, this
converts UV $2S\to1S$ Lyman photons into the visual continuum siding
with the visible Balmer series~$X\to 2S$.  Higher-order processes fall
into still more remote spectral windows, while our leapfrog processes,
even from three or higher-number of photons, are all over the place
along with the single photons that one sees in the spectral shape. The
second variation is our reliance on the quantum-optical
quantity~$g^{(2)}_\Gamma(\omega_1,\omega_2)$ (this would be~$g^{(n)}$
for higher photon numbers), which allows to ``reveal'' this hitherto
unsuspected structure, whose existence is, again, not conditioned to
the amount of signal, and can be captured by using an adequate
two-photon observable, which, as Glauber clarified for quantum light,
is the intensity-intensity correlator~\cite{glauber63a}.  As a result,
one should look for such multiphoton emission \emph{everywhere}, not
only when it is betrayed, for one reason or another, at the one-photon
level. In most cases, it remains invisible (at the one photon
level). The appropriate photon correlations allow us to extract them
from a background of unrelated or more intense emission.  Another
variation is that, two-photon light being more complex than
single-photon one, it comes with more characteristics, that we can
capture with colors as opposed to a magnitude. We use white for
uncorrelated emission where the two photons are emitted independently
the one from the other. Interestingly, one can see in
Fig.~\ref{fig:Sun7Jan183134CET2024} that such uncorrelated photons
form a grid of horizontal and vertical lines defined by the (one
photon) spectral peaks. This means that photons from the spectrum
contribute not only the bulk, indeed, of the emission, but more
significantly, the classical signal which is independent from the rest
of the emission. This is another call for quantum opticians to seek
their photons away from the peaks: quantum photons are emitted where
one does not seem them. This is because photons from the spectral
peaks are ``real'' photons that arise from the Bohrian jump from one
real (here, dressed) state to another, while leapfrog photons are
virtual, in the sense that they involve a virtual intermediate state
with energy~$E_l^*$ which could be any value in between, and thus
resulting in a correlated signal for the entire line satisfying
Eq.~(\ref{eq:Sun7Jan225757CET2024}).  Virtual does not mean
``non-existing'' and would one still complain of no emission outside
from the peaks, we would retort again that this is a one-photon
concern. The no-emission at the two-photon level is unrelated to
no-emission at the one-photon one. No two-photon emission is in fact
what we encode with the blue color in the two-photon spectrum. And one
can see in Fig.~\ref{fig:Sun7Jan183134CET2024}(d) how the strongest no
two-photon emission occurs at $(\varpi_1,\varpi_2)=(-1,-1)$
or~$(+1,+1)$, i.e., at the spectral side peaks, where one-photon
emission is indeed strong. Such a scenario is well known, as it
corresponds to single-photon emission. In contrast, at
$(\varpi_1,\varpi_2)={1\over2}(-1,-1)$ or ${1\over2}(+1,+1)$, one has
the opposite situation of strong two-photon emission but small
one-photon emission (in fact the smallest emission in the range
{$|\varpi|\lessapprox 1.25$}), which corresponds to the equally
fundamental case of two-photon leaprog emission, which may be even
more important than single-photon emission, although, because it is
innocuous at the one-photon level, it failed to attract much attention
so far. One must highlight in this regard the exceptional
contributions from the beautiful and still unique experiments of the
Muller group~\cite{peiris15a,peiris17a,nieves18a,nieves20a}, who
has observed these features in spectacular agreement with the
theory. Finally, a manifestation of the independent yet tangible
existence of the two-photon physics can be based on arguments of
theoretical aesthetics: the robust spectral (one-photon) symmetry,
which appears to be lifted at the two-photon level since the two peaks
behave differently in the two-photon spectrum, is instead ``rotated'':
the symmetry is with respect to the two-photon diagonal, so
independent from the one-photon structure, but also present when
looked at properly. This needed change of perspective, we believe, is
an evidence of the independent two-photon picture.  If such arguments
fail to move one's sensibility, and although we are focused on the
fundamental aspects in this text, we should then mention that there
are obvious and immediate technological prospects of these results
that turn these correlated virtual photons into quantum emission of a
new type. For instance, placing a cavity at the $N$-photon leapfrog
degenerate frequency (halfway between the peaks for two photons) would
Purcell-enhance their emission and open a bright channel of pure
$N$-photon emission~\cite{sanchezmunoz14a}. This is clearly another
evidence of the ``existence'' of such processes, that can power
devices of a new type. There are other ways to similarly exploit this
hidden physics, but we wish to return instead to its basic structure,
and now focus on the perplexing blue circle cast between the central
and virtual peaks, as seen in
Fig.~\ref{fig:Sun7Jan183134CET2024}(f). This circle is of a different
nature from the phenomenology that we have discussed so far, since
this is a curve as opposed to straight lines, and suppressing photons
instead of having them come together as joint emission. We now explain
here for the first time the underlying mechanism for this circle.

\section{Admixing two-mode squeezed and coherent states}
\label{sec:Sun7Jan155552CET2024}

Two-photon physics is primarily described by Fock states~$\ket{2}$ in
a given mode, or by non-degenerate pairs~$\ket{1_11_2}$ in two
modes. Next comes the so-called \emph{squeezed states}, which was
popularized by Walls~\cite{walls83a} for their ability to squeeze
through the uncertainty principle~\cite{caves81a}, but were initially
(and possibly, more appropriately) called ``two-photon coherent
states'' by Yuen~\cite{yuen76a}.  Because two-photon spectra correlate
two frequencies (providing two continuously varying modes), two-mode
squeezing can be expected to have some relevance in presence of
coherence, as is the case for resonance fluorescence.  Two-mode
squeezing considers two bosonic modes with annihilation
operators~$a_{1}$ and~$a_2$, and a so-called ``squeezing
matrix''~$\zeta_{ij}$ (complex-valued symmetric). The two-mode
squeezing operator~\cite{ma90a}
\begin{equation}
  \label{eq:Tue6Feb130406CET2024}
  \mathcal{S}_2^{(ij)} = \exp[ \zeta_{ij} a_i a_j - \zeta_{ij}^* \ud{a_i} \ud{a_j}],
\end{equation}	
can be seen as a generalization of the already introduced single-mode
squeezing operator
\begin{equation}
  \label{eq:Tue6Feb130453CET2024}
  \mathcal{S}_1^{(i)} = \exp[\frac{1}{2} ( \xi_i^* a_i^2 - \xi_i \ud{a_i}^{2} ) ]\,,
\end{equation}
where the squeezing parameters are conveniently defined as
$\xi_i = r_i e^{i \theta_i}$ for~$i=1, 2$ and
$\zeta_{ij} = t_{ij} e^{i \vartheta_{ij}} $ a matrix with zero
diagonal elements, namely $\zeta_{12} = t_{12} e^{i \vartheta_{12}}$
with~$\zeta_{jj}=0$ for~$j=1,2$. One can squeeze the two modes
independently, with the product of the single-mode squeezing operators
$\mathcal{S}_1^{(1)} (\xi_1) \mathcal{S}_1^{(2)} (\xi_2) =
\mathcal{S}_1^{(2)} (\xi_2) \mathcal{S}_1^{(1)} (\xi_1)$ (the
operators commute), or squeeze them jointly with the two-mode
squeezing operator $\mathcal{S}_2^{(12)} (\zeta_{12})$, transforming
annihilation operators as~\cite{ma90a}:
\begin{equation}
  \label{eq:Thu20Oct190623CEST2022}
  (\mathcal{S}_2^{(12)})^\dagger a_i \mathcal{S}_2^{(12)} = \sum_{k=1,2}
  \big(\mathcal{M}_{ik} a_k - \mathcal{N}_{ik} \ud{a_k} \big) \,,
\end{equation}
for~$1\le i\le 2$.  The mixing matrix $\mathcal{M}$ is positive while
the second mixing matrix $\mathcal{N}$ can be complex-valued.  For
single-mode squeezing, they are diagonal with values
$\mathcal{M}_{ii} = \cosh(r_i)$ and
$\mathcal{N}_{ii} = e^{i \theta_i}\sinh(r_i)$, respectively.
On the other hand, for two-mode squeezing, the non-vanishing
elements are $\mathcal{M}_{11} = \mathcal{M}_{22} = \cosh(t_{12})$ and
$\mathcal{N}_{12} = \mathcal{N}_{21} = e^{i
  \vartheta_{12}}\sinh(t_{12})$.  This leads to the transformation
rules for the one- and two-mode operators:
\begin{equation}
  \label{eq:Sun7Jan135507CET2024}
  (\mathcal{S}_1^{(i)})^\dagger a_i \mathcal{S}_1^{(i)} = \ \mu_i a_i - \nu_i \ud{a_i} \quad\text{and}\quad
  (\mathcal{S}_2^{(12)})^\dagger a_i \mathcal{S}_2^{(12)} =  \ \mathcal{M}_{11} a_i - \mathcal{N}_{12} \ud{a_{\bar\imath}} \,,
\end{equation}
where $\mu_i = \cosh(r_i)$, $\nu_i = e^{i \theta_i} \sinh(r_i)$,
$\mathcal{M}_{11}= \cosh(t_{12})$,
$\mathcal{N}_{12}= e^{i \vartheta_{12}}\sinh(t_{12})$ and
$\bar\imath \equiv 3 - i$ exchanges $1$ and $2$.  Alternatively, one
could use the general form of $\mathcal{S}_2^{(12)}$, with
structurally identical results.  The combination of these squeezing
leads to a generic two-mode squeezed state:
\begin{equation}
  \label{eq:Thu20Oct180902CEST2022}
  \ket{\xi_1, \xi_2, \zeta_{12}} \equiv \mathcal{S}_1^{(1)} \mathcal{S}_1^{(2)} \mathcal{S}_2^{(12)} \ket{0 \, 0}.  
\end{equation}
For this state, it is straightforward to compute any correlator
$\av{\ud{a_1}^m a_1^n \ud{a_2}^{p} a_2^q}$ for integers~$m, n, p, q$
from Eqs.~(\ref{eq:Thu20Oct190623CEST2022}). For instance, we find for
the population of mode~$i=1,2$:
\begin{equation}
  \label{eq:Thu20Oct183858CEST2022}
  \av{\ud{a_i} a_i} = \cosh^2 r_i \sinh^2 t + \sinh^2 r_i \cosh^2 t \,.
\end{equation}
One could carry-on like this and compute other correlators but since
we shall be concerned in the following with leading-order processes,
we can rescale the squeezing parameters as
$r_i \rightarrow \epsilon^2 \, r_i$ and
$t \rightarrow \epsilon^2 \, t$ for an $\epsilon$ that will be taken
close to zero. In this limit, thanks to normalization, we can give a
more comprehensive list of the correlators, including the second-order
correlation functions
$g^{(2)}_i\equiv{\av{\ud{a_i}^2 a_i^2}}/{\av{\ud{a_i}a_i}^2}$
for~$i=1,2$ and
$g^{(2)}_{12}\equiv{\av{\ud{a_1}\ud{a_2} a_2 a_1}}/{\av{\ud{a_1} a_1}
  \av{\ud{a_2} a_2}}$:
\begin{subequations}
  \label{eq:Thu20Oct184230CEST2022}
  \begin{align}
    &\av{\ud{a_i} a_i} \approx   \left( r_i^2 +t_{12}^2\right) \epsilon^4\,, \qquad
      \av{a_1a_2} \approx - \epsilon^2  \zeta_{12}\,,  \qquad\av{a_i^2}\approx-\epsilon^2\xi_i\,,\\
    &g^{(2)}_i \approx 
\frac{r_1^4}{(r_1^2 + t_{12}^2)(r_2^2 + t_{12}^2) \epsilon^4} \,, \qquad
    g^{(2)}_{12} \approx 
      \frac{t_{12}^4}{(r_1^2 + t_{12}^2)(r_2^2 + t_{12}^2) \epsilon^4} \,.
  \end{align}
\end{subequations}
Since~$\epsilon$ is very small, one can see that all the two-particle
fluctuations are bunched. We can, however, produce antibunching by
interfering squeezing with a coherent state, as we already discussed
for the Heitler regime at resonance.  Here, we will look more closely
at the general case of interferences that involve two-mode squeezing,
and show that this captures much of the two-photon physics of
resonance fluorescence.  This extends to two modes the idea already
scrutinized for one
mode~\cite{lopezcarreno18b,zubizarretacasalengua20a}, that admixture
of coherence with squeezing results in rich and qualitatively
different correlations than those available in the squeezed or
coherent states alone. This should similarly allow to generalize the
specific technique of tuning photon statistics with coherent
fields~\cite{zubizarretacasalengua20b} to multimode correlations,
especially as multimode coherent states are not correlated, so they
can be tuned independently for each squeezed modes.  From the weak
driving that defines resonance fluorescence, it is enough to deal with
coherent-squeezed states in the limit of small squeezing.  Since the
coherent contribution of the quantum states is taken of the same order
than squeezing, we take $\alpha_i \rightarrow \epsilon \, \alpha_i$,
meaning, however, that coherence is stronger since it appears to first
order in the admixture while squeezing is of second order. This brings
us to the fundamental object of two-photon resonance fluorescence:
\begin{equation}
  \label{eq:Thu20Oct194043CEST2022}
  \ket{\bm{\alpha}, \bm{\xi}, \zeta_{12}} = \mathscr{D}_1 \mathscr{D}_2 \,
  \mathcal{S}_1^{(1)} \mathcal{S}_1^{(2)} \mathcal{S}_2^{(12)} \ket{0 \, 0}\,,
\end{equation}
where~$\bm{\alpha}\equiv(\alpha_1,\alpha_2)$
and~$\bm{\xi}\equiv(\xi_1,\xi_2)$ admixes coherence and squeezing of
single modes~\cite{zubizarretacasalengua20b} along with~$\zeta_{12}$
that provides the general two-photon physics. Similarly as before, we
can now compute the key two-photon correlators for this state, to
leading order in~$\epsilon$:
\begin{subequations}
  \label{eq:Thu20Oct200354CEST2022}
  \begin{align}
  \phantom{a}\hskip-2cm  \av{\ud{a_i}a_i} \approx & \ \epsilon^2 |\alpha_i|^2 + \epsilon^4 \big\{ t_{12}^2 + r_i^2 -  2 |\alpha_i|^2 r_i \cos(2\phi_i - \theta_i) - \ 2 |\alpha_1| |\alpha_2| t_{12} \cos(\phi_1 + \phi_2 - \vartheta_{12}) \big \}  \,, \label{eq:Thu20Oct231212CEST2022}\\
    \av{a_i^2} \approx & \ \epsilon^2 \left(\alpha_i^2- \xi_i  \right) \,, \quad
                         \av{a_1 a_2} \approx  \ \epsilon^2 \left( \alpha_1 \alpha_2 - \zeta_{12}\right) \,, \\                         
\g{2}_i \approx & 1 - \frac{2 \, r_i \cos(2 \phi_i -\theta_i)}{|\alpha_i|^2} + \frac{r_i^2}{|\alpha_i|^4} \,, \\
\g{2}_{12} \approx & \  
1 - \frac{2 \, t  \cos(\phi_1 + \phi_2 - \vartheta_{12})}{|\alpha_1| |\alpha_2|}  +\frac{t_{12}^2}{|\alpha_1|^2 |\alpha_2|^2} \,,\label{eq:Tue6Feb203310CET2024}
  \end{align}
\end{subequations}
for~$1\le i,j\le 2$. As compared to
Eqs.~(\ref{eq:Thu20Oct184230CEST2022}), one can see that the
correlators~$g^{(2)}$ can now take a much wider span of possible
values, in particular they can now be less than unity and even vanish
exactly as well as diverge to leading order. This is the two-mode
generalization of our previous single-mode
admixing~\cite{zubizarretacasalengua20b}, where it was also the case
that the population, Eq.~(\ref{eq:Thu20Oct231212CEST2022}), is
essentially coherent, since~$\epsilon^2\gg\epsilon^4$, but that,
regardless of this preponderance of the coherent state, two-photon
observables are ruled by the squeezing component.  Here too, the
intensity of the squeezed photons has to be much smaller than the
coherent one. The reason is that in order to interfere at the
two-photon level (to produce antibunching) the contributions have to
be of the same order, what we have achieved thanks to the
parametrization.  The coherent field makes the system gain one photon,
with a probability proportional to $|\alpha_i|$. To climb up from
vacuum to $\ket{2}$, the system has to absorb two photons, one each
time, so the chances are proportional to $|\alpha_i|^2$. On the other
hand, the squeezed field carries the photons two by two. Then, the
system can directly jump from $\ket{0}$ to $\ket{2}$ by absorbing two
photons with a probability proportional to $r_i$. To ensure that both
processes compete on equal footing, the ratio $|\alpha_i|^2 / r_i$ has
to remain finite when the limits $|\alpha_i|, r_i \rightarrow 0$ are
taken. For the two-mode case, the interference yields the cancellation
of the $\ket{1,1}$ state and the argument is exactly the same just
exchanging $|\alpha_i|^2 \rightarrow |\alpha_1||\alpha_2|$ and
$r_i \rightarrow t_{12}$. This explains the counter-intuitive fact
that a weak incoherent fraction determines the antibunching of an
intense coherent radiation~\cite{lopezcarreno18a,hanschke20a}.  A
major departure from two-mode squeezing as compared to single-mode
admixing, is the degeneracy for the conditions that realize
interferences producing extreme correlations, such as perfect
antibunching (exact zero) or diverging superbunching. Indeed, perfect
anticorrelation for a single mode is obtained when these conditions
are fulfilled:
\begin{align}
  \label{eq:Thu20Oct232200CEST2022}
  r_i = |\alpha_i|^2\,, && 2 \phi_i - \theta_i = 0\,,
\end{align}
while two-mode anticorrelations are maximized when:
\begin{align}
  \label{eq:Thu20Oct232206CEST2022}
  t_{12} = |\alpha_1\alpha_2|\,, && \phi_1 + \phi_2 - \vartheta_{12} = 0\,.
\end{align}
Just like two-photon leapfrog transitions
(cf.~Sec.~\ref{sec:Sun7Jan122947CET2024}) lift the strict conservation
of energy for each photon to constrain their sum instead---allowing
each photon in the pair to come with a continuous
energy~\cite{gonzaleztudela13a,lopezcarreno17a}---here, the
wave-interference is realized for a sum of the phases, in contrast to
the single phase for a single mode~\cite{zubizarretacasalengua20a}.
This allows, again, for a continuous range of parameters to realize
the quantum interference. Such admixtures can be tracked in any
dynamical two-mode quantum state, possibly mixed with a density matrix
$\rho$. We will consider, explicitly, the case of resonance
fluorescence, which is the simplest one, but of course our discussion
is general, whether the coherence is provided internally (as is the
case in resonance fluorescence where it is inherited from the coherent
driving) or is explicitly added externally, with a supplementary
(homodyning) laser, a case we shall also touch upon. It could also be
developed by the system itself (e.g., when it undergoes
lasing). Whatever the origin for the various contributions, one can
separate them from the whole operators~$a_i$ that act on the whole
state from those~$\tilde a_i$ that act only on the fluctuations, or
incoherent part. They are linked by the relation
\begin{equation}
  \label{eq:Thu20Oct235517CEST2022}
  a_i=\alpha_i+\tilde a_i
\end{equation}
where, again, $a_i$ and~$\tilde a_i$ are operators while $\alpha_i$ is
a $c$-number. This again generalizes our earlier
case~\cite{zubizarretacasalengua20b} and while we limit ourselves here
to~$i=1,2$, one could further extend it to any number of modes. By
construction, one has $\langle\tilde a_i\rangle=0$, resulting in no
interferences for the intensities:
$\langle\ud{a_i}a_i\rangle=|\alpha_i|^2+\langle\ud{\tilde a_i}\tilde
a_i\rangle$. In stark contrast, the cross-correlator
$\av{\ud{a_1}\ud{a_2}a_2a_1}$ becomes
$\av{(\ud{\tilde a_1} + \alpha_1^* ) (\ud{\tilde a_2} + \alpha_2^*)
  (a_2 + \alpha_2) (a_1 + \alpha_1)}$. Expanding this, we get:
\begin{multline}
\label{eq:Fri21Oct000253CEST2022}
\av{\ud{a_1} \ud{a_2}a_2 a_1} = 
 \av{\ud{\tilde a_1} \ud{\tilde a_2} \tilde a_2 \tilde a_1} 
+
\sum_{i=1,2}\big(\alpha_i\av{\ud{\tilde a_1} \ud{\tilde a_2} \tilde a_{\bar\imath}}+\mathrm{cc}\big)+{}\\
+\big(\alpha_1\alpha_2\av{\ud{\tilde a_1}\ud{\tilde a_2}}+\alpha_1\alpha_2^*\av{\ud{\tilde a_1}{\tilde a_2}}+\mathrm{cc}\big)
+\sum_{i=1,2}|\alpha_i|^2\av{\ud{\tilde a_{\bar\imath}}\tilde a_{\bar\imath}}
\end{multline}
where~$\bar\imath\equiv 3-i$, i.e., toggles the index in~$\{1,2\}$.
This shows that, unlike intensities, the second-order correlation
functions are the result of interferences, namely, of four terms:
\begin{equation}
  \label{eq:Fri21Oct003956CEST2022}
  g^{(2)}_{12}=1 + \mathcal{I}_0 +  \mathcal{I}_1 + \mathcal{I}_2\,,
\end{equation}
with
\begin{subequations}
  \label{eq:Fri21Oct004220CEST2022}
  \begin{align}
    \mathcal{I}_0 & \equiv{1\over n_1n_2}\left\{ \av{\ud{\tilde a_1} \ud{\tilde a_2} \tilde a_2 \tilde a_1} - \av{\ud{\tilde a_1} \tilde a_1} \av{\ud{\tilde a_2} \tilde a_2} \right\}\, , \\
    \mathcal{I}_1 & \equiv {2\over n_1n_2}\operatorname{Re}\left\{{\alpha_1 \av{\ud{\tilde a_1} \ud{\tilde a_2} \tilde a_2} + \alpha_2 \av{\ud{\tilde a_1} \ud{\tilde a_2} \tilde a_1} }\right\} \, , \\
    \mathcal{I}_2 & \equiv {2\over n_1n_2}\operatorname{Re}\left\{\alpha_1 \alpha_2 \av{\ud{\tilde a_1} \ud{\tilde a_2}} + \alpha_1 \alpha_2^* \av{\ud{\tilde a_1}  \tilde a_2} \right\}\, , 
  \end{align}
\end{subequations}
with~$n_i\equiv\av{\ud{a_i}a_i}$ so that in each case,
$\mathcal{I}_in_1n_2$ is of order~$|\alpha_1|^p|\alpha_2|^q$
with~$p+q=i$. We now have recovered the general version of the
phenomenon discussed in Section~\ref{sec:Sat10Feb185005CET2024} of
antibunching produced by an interference of various contributions,
which are the single-mode $\mathcal{I}_i$, i.e., with~$a_1=a_2=a$, in
which case~$\mathcal{I}_0$ is related to the $g^{(2)}$ of the
incoherent fraction, being $-1$ for non-Gaussian (quantum) antibunching
and~$+1$ for chaotic, bunched states, while~$\mathcal{I}_2$ is related
to squeezing, being~$-2$ in presence of squeezing.  $\mathcal{I}_1$ is
a measure of anomalous moments which is negligible in the two regimes
of interest, namely
\begin{itemize}
\item Heitler regime, where~$g^{(2)}=0=1+(\mathcal{I}_0=1)+(\mathcal{I}_1=0)+(\mathcal{I}_2=-2)$,
\item Mollow regime,  where~$g^{(2)}=0=1+(\mathcal{I}_0=-1)+(\mathcal{I}_1=0)+(\mathcal{I}_2=0)$,
\end{itemize}
with, therefore, a completely different interpretation for how
$g^{(2)}=0$ is obtained.  In the general case that we consider here,
the numerator of $\mathcal{I}_1$ has the physical meaning of a
covariance between the incoherent fractions, being positive when both
modes are occupied or depleted together, and negative when one is
largely occupied and the other depleted. As for the single-mode, the
normalization is to the \emph{total} population, not to that of the
incoherent fraction alone. $\mathcal{I}_2$ also has some
interpretation in terms of two-mode squeezing, in particular including
the coherent fractions~$\alpha_1$ and~$\alpha_2$. We shall similarly
disregard the exact meaning of~$\mathcal{I}_1$ that interpolates
between the two cases by involving anomalous correlators of the
population and order parameter of the
type~$\langle a_i\hat n_{\bar\imath}\rangle$, also weighted by the
coherent fraction $\alpha_i^*$.  We can also write these expressions
not for the quantum fields only (fluctuations), but for the full-state
correlators, which might be more accessible either experimentally or
theoretically. To do so, we can use the backward relation:
\begin{multline}
  \label{eq:Sun7Jan174938CET2024}
  \av{\tilde a_1^{\dagger i} \tilde a_2^{\dagger j} \tilde a_2^k \tilde a_1^l} =  
  \sum_{i', j', k', l' = 0}^{i,j,k,l} (-1)^{i'+j'+k'+l'}
  \binom{i}{i'} \binom{j}{j'}
  \binom{k}{k'} \binom{l}{l'} (\alpha_1^{*})^{i-i'} (\alpha_2^{*})^{j-j'}
  \alpha_2^{k'}\alpha_1^{l'} \times \\ \av{ a_1^{\dagger (i-i')} a_2^{\dagger (j-j')} a_1^{k-k'} a_2^{l-l'}},
\end{multline}
which applied to $\mathcal{I}_i$ leads to:
\begin{subequations}
\label{Sun7Jan180413CET2024}	
\begin{align}
  \mathcal{I}_0  = \, & 
                        \big( \av{\ud{a_1} \ud{a_2} a_2 a_1} - \av{\ud{a_1} a_1} \av{\ud{a_2} a_2} - 4 \, |\alpha_1|^2 |\alpha_2|^2 + 2 \, |\alpha_1|^2 \av{\ud{a_2} a_2} + \nonumber \\
                      & 2 \, |\alpha_2|^2 \av{\ud{a_1} a_1} \, + 
                        2 \, \operatorname{Re} \lbrace \alpha_1 \alpha_2 \av{\ud{a_1} \ud{a_2}} + \alpha_1 \alpha_2^* \av{\ud{a_1} a_2}
                        - \alpha_1  \av{\ud{a_1} \ud{a_2} a_2} - \alpha_2  \av{\ud{a_1} \ud{a_2} a_1} \rbrace  \big) /
                        n_1 n_2 ,  \\
  \mathcal{I}_1  = & 2 \, 
                     \operatorname{Re}\big({\alpha_1 \av{\ud{a_1} \ud{a_2} a_2} + \alpha_2 \av{\ud{a_1} \ud{b} a_1} - 2 \, \alpha_1 \alpha_2 \av{\ud{a_1} \ud{a_2}} - 2 \, \alpha_1 \alpha_2^* \av{\ud{a_1} a_2}} + \nonumber \\
                      & \ 2 \, |\alpha_1|^2 |\alpha_2|^2 - |\alpha_1|^2 \av{\ud{a_2} a_2} - |\alpha_2|^2 \av{\ud{a_1} a_1}  \big) /
                        n_1 n_2 , \\
  \mathcal{I}_2 = & 2 \, 
                    \operatorname{Re}\big({\alpha_1 \alpha_2 \, \av{\ud{a_1} \ud{a_2}} + \alpha_1 \alpha_2^* \av{\ud{a_1} a_2} }
                    - 2 \, |\alpha_1|^2 |\alpha_2|^2 \big)  /
                    n_1 n_2\,.
\end{align}
\end{subequations}

One can recover from this general result the particular
case~(\ref{eq:Thu20Oct194043CEST2022}) of a coherent squeezed state by
checking that these quantities
for~$\ket{\bm{\alpha}, \bm{\xi}, \zeta_{12}}$ with vanishing
amplitudes ($\alpha_i \rightarrow \epsilon \alpha_i$
$r_i \rightarrow \epsilon^2 r_i$ and
$t_{12} \rightarrow \epsilon^2 t_{12}$, with
$\epsilon \rightarrow 0$), read
\begin{equation}
  \label{eq:Tue20Feb154414CET2024}
  \mathcal{I}_0 \approx \frac{t_{12}^2}{|\alpha_1|^2|\alpha_2|^2} \,, \qquad \mathcal{I}_2 \approx - \frac{2 t_{12} \cos(\phi_1 + \phi_2 - \vartheta_{12})}{|\alpha_1||\alpha_2|} \,,
\end{equation}
while the anomalous $\mathcal{I}_1$ exactly zero for any set of
parameters. This recovers Eq.~(\ref{eq:Tue6Feb203310CET2024}) from
Eq.~(\ref{eq:Fri21Oct003956CEST2022}) and, with the conditions for the
cancellation of $\g{2}_{12}$ in Eq.~\eqref{eq:Thu20Oct232206CEST2022},
this gives
\begin{equation}
  \label{eq:cancellation2modeIterms}
  \mathcal{I}_0 \approx 1 \,, \qquad \mathcal{I}_2 \approx - 2 \,,
\end{equation}
i.e., the compensation between the mean field and the squeezed
two-mode fluctuations that leads to $\g{2}_{12} = 0$ at first order in
the parameters and that we also find in the single-mode
antibunching~\cite{zubizarretacasalengua20b,hanschke20a}.  All the
previous considerations would be interesting, but formal results only,
would it not be for the link that photodetection offers to the theory
of frequency-resolved photon correlations, as discussed in
Section~\ref{sec:Sun7Jan122947CET2024}. Namely, retaining the
frequency information of the photons that one correlates, brings to
the fore the decomposition Eq.~(\ref{eq:Fri21Oct003956CEST2022}) but
now for the continuously varying frequencies~$\omega_i$ as opposed to
independent modes~$a_i$.  We now turn to such a dynamical setting.

\section{Squeezed cavity}
\label{sec:Sun7Jan163313CET2024}

\begin{figure}[htb]
  \centering
  \includegraphics[width=\linewidth]{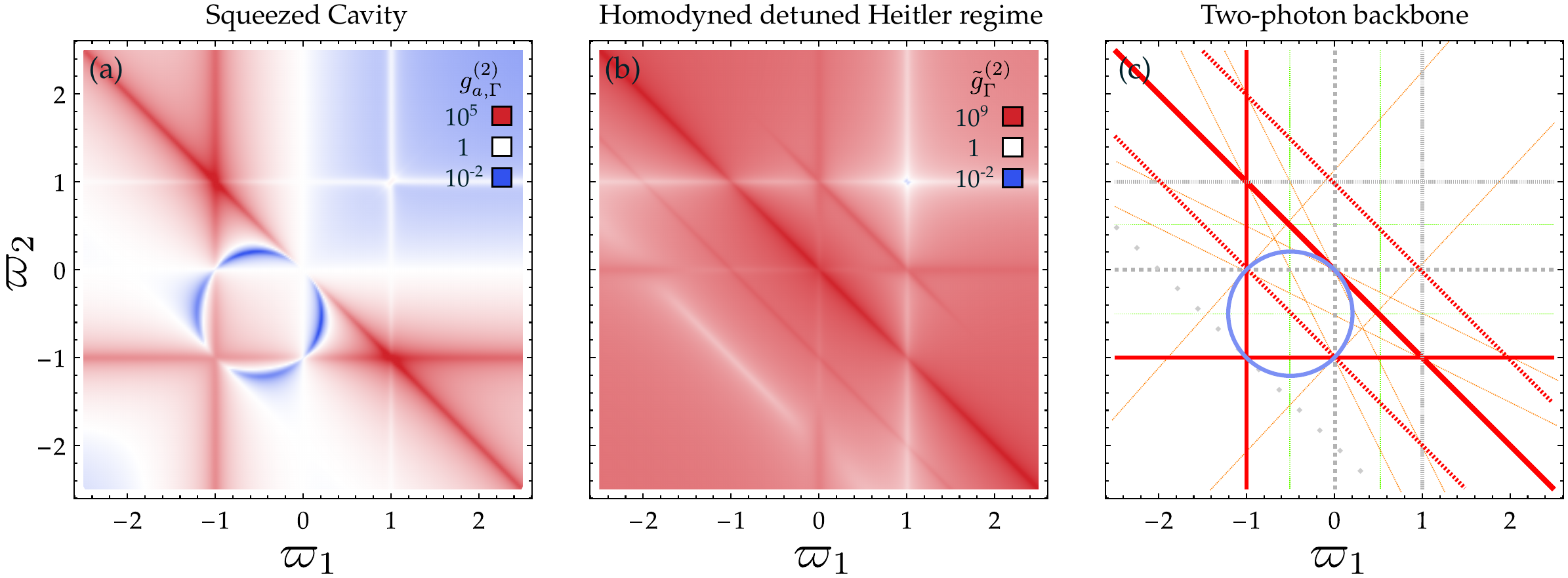}
  \caption{Three manifestations of the origin of the circles of
    antibunching: (a) The numerical two-photon spectrum for the
    squeezed cavity reproduces an almost identical structure to that
    of exact resonance fluorescence
    (cf.~Fig.~\ref{fig:Sun7Jan183134CET2024}(f)). (b) the circle is
    removed when the coherent fraction itself is removed (by
    homodyning), leaving only the strong superbunching leapfrog
    processes.  (c) The analytical expression also provides exact
    straight lines and a circle appearing as
    Eqs.~(\ref{eq:Sun7Jan195928CET2024}--\ref{eq:Wed7Feb111608CET2024})
    and Eq.~(\ref{eq:Sun7Jan193611CET2024}), respectively. The mesh of
    thin lines are those observed in some two-photon landscapes that
    go beyond the physics covered in this text.  Parameters: (a)
    $\gamma_a = 1$, $\Delta_a = 80.1$, $\lambda_a = 0.001$ and
    $\Gamma = 2$, optimizing antibunching. Panel~(b) is the same as
    Fig.~\ref{fig:Sun7Jan183134CET2024}(f) but with an external
    homodyning field~$\alpha=-\langle\sigma\rangle$. }
  \label{fig:Sun11Feb202132CET2024}
\end{figure}

We now return to the two-photon spectrum, to reveal how the concepts
of squeezed and coherent admixtures introduced in
Section~\ref{sec:Sun7Jan155552CET2024} explain the results we reported
for resonance fluorescence, as shown in
Fig.~\ref{fig:Sun7Jan183134CET2024}. We can best show that this is the
case by bringing directly all these components externally to a
cavity~$a$ and collect the re-emitted admixed light, instead of a
two-level system producing them internally. This approach also invites
closer inspection of such excitation schemes~\cite{georgiades99a}.
The sensor technique for this scenario produces the Hamiltonian:
\begin{equation}
  \label{eq:Sun7Jan165826CET2024}
  H_a = \Delta_a \, \ud{a}a + i \frac{\Lambda_a}{2} (a^{\dagger 2} - a^2) + \Omega_a (e^{i \vartheta} \ud{a} + e^{-i \vartheta} a) + \Delta_1 \ud{\varsigma_1} \varsigma_1 + \Delta_2 \ud{\varsigma_2} \varsigma_2  + \epsilon \sum_{i = 1,2} (\ud{a} \varsigma_i + \ud{\varsigma_i} a ) \,.
\end{equation}
Here we have a passive, linear cavity~$a$, driven by both a squeezed
source with strength~$\Lambda_a$ and a laser with
amplitude~$\Omega_{a}$, with $\vartheta$ is the phase difference
between the coherent and squeezed sources. In resonance fluorescence,
only the coherent state is provided externally while the weak
nonlinearity, given the weak driving, produces the squeezing. In other
cases, everything could be produced by the system itself (e.g., like
the Mollow triplet produced under incoherent pumping when placed in a
cavity that undergoes lasing~\cite{delvalle10d}). While the 2LS is
inherently antibunched ($\g{2}_\sigma (0) = 0$), independent of the
nature and strength of the driving, a squeezed coherent cavity may
modulate the photon statistics from bunched to antibunched as the
parameters change. Since we are interested in showing the similarity
of this system with resonance fluorescence, we first seek to find the
conditions that minimize the cavity's two-photon correlator~$\g{2}_a$.
In the steady-state, the bare correlations of the squeezed cavity are
\begin{equation}
  \label{eq:Sun7Jan170545CET2024}
    \mean{a} =  -\frac{2 i \Omega_a [e^{i \vartheta} (\gamma_a - 2 i \Delta_a) - 2 e^{-i \vartheta} \Lambda_a]}{\gamma_a^2 + 4 \Delta_a^2 - 4 \Lambda_a^2} \quad\text{and}\quad
    \langle\ud{a}a\rangle=\frac{2 \Lambda_a ^2}{\gamma _a^2+4 \Delta _a^2-4 \Lambda_a ^2} + |\av{a}|^2\,.
\end{equation}
We do not provide the expression for the general~$\g{2}_a$ as it is
too voluminous and not immediately needed for our discussion, unlike
the variance
$\mean{a^2} - \mean{a}^2=\Lambda_a(\gamma_a - 2 i \Delta_a)/[\gamma_a^2 + 4
\Delta_a^2 - 4 \Lambda_a^2]$
that is required for the phase matching $\theta = 2\phi$ between
$\phi \equiv \arg( \mean{a})$ and
$\theta \equiv \arg(\mean{a^2} - \mean{a}^2)$ needed to
minimize~$\g{2}_a$. This is satisfied
for~$\tan(2 \vartheta) = {2\Delta_a/\gamma_a}$. There is an upper-bound
for the squeezing amplitude~$\Lambda_a$ beyond which the system is
unstable and results in unphysical steady states, namely, it must be
that $\Lambda_a < \sqrt{\gamma_a^2 + 4 \Delta_a^2}\,/2$.
Reparameterizing $\Lambda_a$ as $\Lambda_a = \Gamma_a\lambda / 2 $
where $\Gamma_a \equiv \sqrt{\gamma_a^2 + 4 \Delta_a^2}$, expresses
this stability condition to $\lambda < 1$. The cavity phase and its
population in the phase-matching, i.e., optimum-antibunching
configuration, are then found as
\begin{equation}
  \label{eq:Mon19Feb122922CET2024}
  \mean{a}= \frac{2 i \Omega_a \sqrt{\gamma_a -2 i \Delta_a}}{(1+\lambda)\Gamma_a^{3}} 
  \quad\text{and}\quad \mean{\ud{a} a} = (1+\lambda)^{-2} \left[\frac{4 \Omega_a^2}{\Gamma_a^2} + \frac{\lambda^2 (1 + \lambda)}{2(1-\lambda)}\right] \,,
\end{equation}
with now a tractable two-photon coincidences phase-matched Glauber
correlator (in the~$\Gamma\to\infty$ limit):
\begin{multline}
  \label{eq:Mon19Feb123132CET2024}
  \g{2}_{}a(0) = \frac{(1- \lambda)^2}{[\Gamma_a^2 \lambda^2(\lambda^2 -1)-8\Omega_a^2 (1-\lambda^2)]^2} \times \\
\big[\Gamma_a^4 \lambda^2 (1+\lambda)^2 (1+2 \lambda^2) + 16 \Omega_2^2 \Gamma_a^2
\lambda (\lambda^2-1)(1- 2 \lambda) + 64 (1-\lambda)^2 \Omega_a^4
 \big]\,.
\end{multline}
Deriving Eq.~(\ref{eq:Mon19Feb123132CET2024}) with respect to
$\Omega_a$, we find the minimum value possible for $\g{2}_a$ of a
squeezed cavity admixed to a coherent state:
\begin{equation}
  \g{2}_{a, \mathrm{min}} = 2 \frac{\lambda (2 - \lambda)}{1+2 \lambda - \lambda^2} \overset{\lambda \rightarrow 0}{\approx} 0 + 4 \lambda \,,
\end{equation}
which is obtained for the ``optimum''
driving~$\Omega_{a, \mathrm{optmin}}$:
\begin{equation}
  \label{eq:Wed21Feb133655CET2024}
  \Omega_{a, \mathrm{optmin}} = \frac{\Gamma_a \sqrt{ \lambda/2}(1+\lambda)}{2(1-\lambda)} \overset{\lambda \rightarrow 0}{\approx}
\frac{\Gamma_a}{2}  \sqrt{ \lambda/2}\,.
\end{equation}
This produces antibunching for the whole range $0 \leq \lambda < 1$
and reaches zero as $\lambda$ gets smaller (for both
$\Lambda_a \rightarrow 0$ and $\Delta_a \rightarrow \infty$, which is
the case of interest).  Applying the optimum antibunching conditions
in the limiting case $\Delta_a \rightarrow \infty$ and, therefore
$\lambda \rightarrow 0$, we get for the spectrum
\begin{align}
  \label{eq:Mon19Feb135535CET2024}
  S_{a,\Gamma}(\omega) \approx 
  |\langle a\rangle|^2\mathscr{L}_\Gamma(\omega)
  +(\av{\ud{a}a}-|\av{a}|^2)\frac{2}{\pi} \frac{\gamma_{12}(\gamma_{11}^2+4 \Delta_a^2) + 4 \Gamma \omega^2}{(\gamma_{11}^2 + 4 \omega^2)^2 + 8 \Delta_a^2 (\gamma_{11}^2 - 4\omega^2 )+ 16 \Delta_a^4}
\end{align}
where we also used the shortcut
$\gamma_{ij}\equiv i\Gamma + j\gamma_a$ but this time
for~$\gamma_a$. We will now show that this matches exactly the
two-level system result. The same occurs for the 
two-photon correlator for the squeezed cavity,
\begin{equation}
  \label{eq:Tue20Feb101401CET2024}
  \g{2}_{a} (\tau) \approx 1 + e^{-\gamma_a\tau} - 2 \, e^{- \gamma_a/2 \, \tau } \cos(2 \Delta_a \tau) \,,
\end{equation}
which is identical at resonance to the expression for the unfiltered
two-level system, cf. Eq.~(\ref{eq:Wed7Feb110037CET2024}). In the Next
Section, we show that the agreement holds also in presence of
detuning.  We conclude this preliminary Section with the numerically
exact two-photon spectrum for the squeezed cavity in the optimum
phase-matching, $\theta=2\phi$, and pumping,
Eq.~(\ref{eq:Wed21Feb133655CET2024}), conditions, which is shown in
Fig.~\ref{fig:Sun11Feb202132CET2024}(a). Its excellent qualitative
agreement for most of the features with resonance fluorescence
involving the two-level system, is obvious
(cf.~Fig.~\ref{fig:Sun7Jan183134CET2024}(f)). This confirms the
central importance of quantum state interferences in general and of
coherent and squeezed states in particular, to account for the
phenomenology of two-photon physics of detuned resonance fluorescence.
This is fully established in the next Section.

\section{Detuned resonance fluorescence}
\label{sec:Mon19Feb221640CET2024}

We can now complete our argument by returning to the two-photon
spectrum of resonance fluorescence (i.e., for the two-level
system~$\sigma$). We will focus on detuning, since this highlights
various features of interest, starting with the spectral separation of
the coherent peak, which stays pinned at the center, from the
incoherent emission, that splits into the symmetric doublet of side
peaks, with a neat interpretation in terms of two-photon scattering as
sketched in Fig.~\ref{fig:Fri9Feb094747CET2024}(c)~\cite{dalibard83a}.
Our goal now is to show that the one and two-photon spectra of the
coherently-squeezed cavity derived in the previous Section, are,
beyond their qualitative agreement with the two-level system, in fact
in quantitative agreement to leading order in the driving.  This will
confirm that in this detuned Heitler regime, and at the two-photon
level, the main role of the two-level system is to bring in a
component of squeezing, and that the observed phenomenology then
follows from two-mode interferences between squeezed and coherent
states, with frequencies spanning over all possible modes of the
system.  The one-photon detuned Heitler spectrum $S_\Gamma(\omega)$
from Eq.~(\ref{eq:Wed14Feb000029CET2024}), with
$\Delta_\sigma\gg\Omega_\sigma$, indeed recovers the squeezed-coherent
cavity spectrum~(\ref{eq:Mon19Feb135535CET2024}) with the
correspondence~${\Omega_\sigma^2\over\Delta_\sigma^2}\to{\Lambda_a\over2|\Delta_a|}$
(which are the populations~$\langle\ud{\sigma}\sigma\rangle$
and~$\langle\ud{a}a\rangle$, respectively).  The same occurs with
two-photon correlations, although the general expression being
awkward, it is not practical to display here their mathematical
identity, which we have, however, checked.  We can, instead, highlight
particular cases of interest, including the insightful tampering of
the photon interferences by filtering out the central peak with a
notch filter~\cite{masters23a}.  In this case, instead of filtering
part of the spectrum and correlating it, one filters out another part
and correlate what remains. Such an observable could be derived using
the frequency-filtered approach described previously, but filtering
out the coherent peak is more expediently achieved by homodyning it
out, admixing the total luminescence with a laser of matching
amplitude~$\alpha$ but with opposite phase. If we parameterize this
field as~$\alpha = -\mathcal{F}\mean{\sigma}$ in terms
of~$0\le\mathcal{F}\le 1$, the two-photon correlations for the
homodyned field are found as:
\begin{equation}
  \label{eq:Tue20Feb100905CET2024}
  g^{(2)}_\mathcal{F}(\tau)=1  + \frac{1}{(1-\mathcal{F})^4} e^{-\gamma_\sigma \tau} -  \frac{2}{(1-\mathcal{F})^2} e^{-(\gamma_\sigma/2)\tau} \cos(\Delta_\sigma \tau)
\end{equation}
where the homodyning is varied from~$\mathcal{F}=0$ (no correction, so
the full emission is taken) in which case
Eq.~(\ref{eq:Tue20Feb101401CET2024}) is recovered, to~$\mathcal{F}=1$,
where the coherent peak is removed completely (so only the incoherent
emission is taken), in which case one observes a simple and smooth
bunching
$g^{(2)}_{\mathcal{F}=1}(\tau)=1 + \frac{\Delta_\sigma^4}{4
  \Omega_\sigma^4} e^{-\gamma_\sigma \tau}$, that can be understood as
the two photons~$\omega_v$ and~$\omega_\sigma$ arriving together. The
presence of the laser thus makes them arrive at different times so
that their coincidence is suppressed, but oscillations ensue as a
result of this time redistribution. The effect can also be understood
as beating due to detuning.  More general spectral filtering brings us
directly to the full two-photon spectrum. In presence of homodyning,
i.e., the two-photon spectrum of the incoherent part alone, transforms
the landscape of correlations from
Fig.~\ref{fig:Sun7Jan183134CET2024}(f) to
\ref{fig:Sun11Feb202132CET2024}(b), i.e., the circle disappears. This
is another proof of its origin from interferences between the
squeezing, that remains and produces extreme bunching (cf. the scale),
and coherence. The result is similar for the squeezed cavity, i.e.,
with~$\Omega_a=0$, but without the side leapfrogs, the central cross
and the antibuching around the real peak (not shown).  The general
two-photon spectrum can be obtained analytically (for both the
two-level system and the squeezed cavity, both at and out-of
resonance, with or without homodyning), but it is too voluminous to be
replicated here. Excellent approximations can, however, be derived in
the the highly-detuned regime
$\Delta_{\sigma} \gg \Omega_{\sigma} \gg \gamma_{\sigma}$, where
emission drops due to the inefficient excitation of the system and
thus enters in the Heitler regime, although the driving is taken much
larger than decay.  As a consequence, the main contribution for the
emitted light is the coherent one.  To leading order, the population
of each sensor~$\langle \ud{\varsigma_i} \varsigma_i \rangle$---that
both provide, from the sensor method, the spectral
shape~$S_\Gamma(\varpi)$---recovers the spectral shape
Eq.~(\ref{eq:Mon19Feb135535CET2024}), featuring a diverging
contribution at the origin that corresponds to the coherent $\delta$
peak, in addition to the symmetric side peaks, which are also revealed
as not Lorentzian:
\begin{equation}
  \label{eq:Mon19Feb173842CET2024}
  S_\Gamma(\varpi) = \frac{\epsilon^2 \Omega_\sigma^2}{\Delta_\sigma^4}{1\over \varpi^2} + \frac{2 \epsilon^2 \Omega_\sigma^4 }{\Gamma \Delta_\sigma^6}{\Gamma + 2 \gamma_\sigma + \Gamma \varpi^2\over(1-\varpi^2)^2}+\text{ higher orders}\,.
\end{equation} 
The two-photon correlation spectrum is similarly obtained in both
cases as:
\begin{equation}
  \label{eq:Tue6Feb192146CET2024}
    g^{(2)}_\Gamma(\varpi_1,\varpi_2)=
    \frac{({\varpi}_1^2 + {\varpi}_1 + {\varpi}_2^2 +{\varpi}_2)^2}
  {({\varpi}_1 + {\varpi}_2)^2({\varpi}_1 + 1)^2 ({\varpi}_2 + 1)^2}
  + \text{higher orders.}
\end{equation}
This surprisingly compact expression gives an excellent qualitative
account of the detuned resonance fluorescence landscape shown in
Fig.~\ref{fig:Sun7Jan183134CET2024}(f), with divergences and exact
zeros that capture the resonances, which get tamed by the higher-order
corrections as is the case for populations with
Eq.~(\ref{eq:Mon19Feb173842CET2024}). From the analysis of the zeros
of both the numerator and denominator of this expression, we can thus
locate the main features of the two-photon spectrum.  Minima
correspond to antibunching. They are found by setting the numerator of
Eq.~(\ref{eq:Tue6Feb192146CET2024}) to zero, which, rewriting the
equation as
\begin{equation}
  \label{eq:Sun7Jan193611CET2024}
  \Big(\varpi_1+{1\over2}\Big)^2+\Big(\varpi_2+{1\over2}\Big)^2={1\over2}  
\end{equation}
is that of a circle centered at $\varpi_1 = \varpi_2 = -1/2$ and with
radius $\sqrt{2}/2$. This not only gives the most compelling
explanation for the circle of antibunching---as a two-mode
interference between squeezing and coherence---it also shows that it
is, indeed, an exact circle.  Bunching maxima are actually divergences
to leading order, and located at the condition that vanish the
denominator,~i.e.,
\begin{equation}
  \label{eq:Sun7Jan195928CET2024}
  {\varpi}_1 = -1\,,\quad {\varpi}_2 = -1\quad\text{and}\quad
  {\varpi}_1 + {\varpi}_2=0
\end{equation}
producing the three main bunching lines of detuned resonance
fluorescence, which make a right-angled triangle.  Additional maxima
can be found looking for the zeros of higher orders (less pronounced
in the plot). These are at
\begin{equation}
  \label{eq:Wed7Feb111608CET2024}
  {\varpi}_1 = 1\,,\quad  \varpi_2=1\quad\text{and}\quad \varpi_1+\varpi_2=\pm1\,.
\end{equation}
The horizontal and vertical correlations now concern the real peak,
which is much less correlated than its virtual counterpart. The
antidiagonal lines are the other leapfrog processes.  All these lines,
along with the circle, are plotted in
Fig.~\ref{fig:Sun7Jan183134CET2024}(c) as a backbone for the
two-photon correlation spectrum.  The decomposition of the 2PS into
its interference terms, using Eqs.~\eqref{Sun7Jan180413CET2024}, leads
to:
\begin{subequations}
  \label{eq:Wed7Feb111704CET2024}
  \begin{align}
    \mathcal{I}_0 & \approx \frac{\varpi_1^2 \varpi_2^2 (2 + \varpi_1 + \varpi_2 )^2}{(\varpi_1 + \varpi_2)^2(\varpi_1 + 1)^2 (\varpi_2 + 1)^2} \,,\label{eq:Mon19Feb142051CET2024} \\
    \mathcal{I}_1 &\approx 0 \,, \\
    \mathcal{I}_2 & \approx -\frac{2\varpi_1 \varpi_2 (2 + \varpi_1 + \varpi_2 )}{(\varpi_1 + \varpi_2)(\varpi_1 + 1)(\varpi_2 + 1)}\,.\label{eq:Mon19Feb142102CET2024}
  \end{align}
\end{subequations}
This reveals that, furthermore
\begin{equation}
  \label{eq:Mon19Feb181955CET2024}
  \mathcal{I}_0=(\mathcal{I}_2/2)^2  
\end{equation}
which, from Eq.~(\ref{eq:Fri21Oct003956CEST2022}), provides~$\g{2}$ as
a function of the squeezing component alone:
\begin{equation}
  \label{eq:Mon19Feb155727CET2024}
  g^{(2)}_\Gamma=\left(1+{\mathcal{I}_2\over2}\right)^2\,.
\end{equation}
This relationship holds when the coherent fraction dominates over the
incoherent, or quantum, fraction, with also phase-matching between
them, in which case, squeezing can be related to photon
fluctuations~\cite{mandel82a,loudon84b}. This is the case for instance
in Eqs.~(\ref{eq:Tue20Feb154414CET2024}) when
$\phi_1+\phi_2-\vartheta_{12}=0$ or~$\pi$ This is also the case in the
Heitler regime where the coherent fraction overtakes the
signal~\cite{lopezcarreno18b,zubizarretacasalengua20a}, including in
presence of filtering~\cite{hanschke20a}. This relationship is the
chief reason why squeezing has been more difficult to observe than
antibunching~\cite{schulte15a}. In more general situations, in
particular when~$\mathcal{I}_1$ is nonzero, there are deviations from
this ideal. The exact numerical results of~$\mathcal{I}_i$ for
resonance fluorescence are shown in
Fig.~\ref{fig:Sun7Jan183134CET2024}(g--o). The rightmost column of the
detuned Heitler regime is well approximated by
Eqs.~(\ref{eq:Wed7Feb111704CET2024}) with, in particular,
$\mathcal{I}_1\approx0$ (the darker area means a negative sign for the
plotted quantity) and $\mathcal{I}_0$ and~$\mathcal{I}_2$ having the
redundant appearance from Eq.~(\ref{eq:Mon19Feb181955CET2024}). The
squeezing component changes sign on its various domains, as shown in
the figure. The condition that produces the circle of exact
antibunching leads to
\begin{equation}
  \mathcal{I}_0 \approx 1 \,, \quad \mathcal{I}_1 \approx 0 \,, \quad
  \mathcal{I}_2 \approx -2 \,,
\end{equation}
which are the same as those for two-photon suppression from
destructive interferences between the squeezing and coherent
component, but here occurring at the two-photon (with two different
frequencies) level.  Since~$\mathcal{I}_2<0$ for this to occur, this
happens in the four triangles that corner the circle in
Fig.~\ref{fig:Sun7Jan183134CET2024}(o). The right-angled triangle is
also interesting. Its lines originate from~$\mathcal{I}_0$, like the
leapfrog processes at resonance. They indeed lie at the frontier of
$\mathcal{I}_2$ changing sign, as can be seen in
Fig.~\ref{fig:Sun7Jan183134CET2024}(o) where they match the dotted
lines where~$\mathcal{I}_2=0$, although it is large (in absolute
value) on both sides. The antidiagonal line corresponds to the central
leapfrog line. The two new lines that appear, horizontal and vertical,
betray the virtual character of the peak, which remains correlated, at
this \emph{pinned} frequency (this is the novelty as compared to other
leapfrog photons), with all the other photons, at all the frequencies,
including the real photon from the other peak, in which case one
realizes the heralded two-photon
scheme~\cite{schrama92a,ulhaq12a,zubizarretacasalengua23a}.  It is
noteworthy that not only the circle, but also the vertical and
horizontal lines disappear with homodyning
(cf.~Fig.~\ref{fig:Sun11Feb202132CET2024}(b)), which suggests that the
anomalous correlator~$\mathcal{I}_1$ plays an important role in the
constitution of these ``anomalous'' lines, which therefore demand
further attention, that is beyond the scope of our current discussion.
In contrast, in the first column of
Fig.~\ref{fig:Sun7Jan183134CET2024}, i.e., at resonance, one has the
opposite situation where~$\mathcal{I}_0\gg\mathcal{I}_2$ and,
furthermore, $\mathcal{I}_0<0$ in the two diagonal quadrants, which
contain the spectral peaks (with a meticulous exclusion of the
leapfrogs). This signals the non-Gaussian, i.e., Fock, or multiphoton
emission, character of both the leapfrogs which have no squeezing
associated to their constitution, and of the side peaks, which have
the characteristic~$\mathcal{I}_0$ of antibunching dominated by the
fluctuations of the incoherent emission. The second column as the
system transits from one case to the other shows the ``evaporation''
of the negative domain in~$\mathcal{I}_0$, which reduces to a tiny
islet around the real peak only at high detuning, as well as the
emergence and shaping of squeezing in~$\mathcal{I}_2$. It also shows
the role of the anomalous correlator~$\mathcal{I}_1$ in bridging
between these two cases, being close to negligible in the other
limits.

This achieves our description of the two-photon landscape of resonance
fluorescence. All the features have been explained. There is a rich
combination of various mechanisms, from multiphoton emission to
multiphoton interferences, being realized in different regions of
phase space where they can be isolated and exploited. We have largely
focused on zero time-delay, but one could similarly consider the time
dynamics of this physics, as we did in
Ref.~\cite{zubizarretacasalengua23a} where, cross-correlating the side
peaks of detuned resonance fluorescence with or without homodyning the
coherent peak, we observed a considerable transformation and
enhancement of the correlation functions. Clearly, the problem is far
from being exhausted. Also it should be clarified that although at the
two-photon level, the physics is captured by two-mode squeezing, the
broader problem including higher photon numbers goes much beyond this
vantage point. One should not make the same mistake with two-photon
spectra as has been made with single-photon ones, i.e., assuming that
the picture is now complete. This might be the case for the squeezed
cavity, since it only involves Gaussian states and thus is fully
described through its two-photon correlators, higher-order ones being
functions of those. But the two-level system gives rise to
non-Gaussian states, and so, the three-photon spectrum (and others of
higher orders) are most certainly providing as radical changes as
compared to the two-photon case than the two-photon case does to the
single-photon one. Some of this higher-order physics does in fact
indeed transpire in two-photon observables.  In the next Section, we
give a brief overview of a few alternative quantifiers of two-photon
emission. There, one will get a chance to sight unaccounted-for
phenomena.

\section{Violation of Inequalities}
\label{sec:Sun11Feb211557CET2024}

\begin{figure}[!h]
  \centering\includegraphics[width=\linewidth]{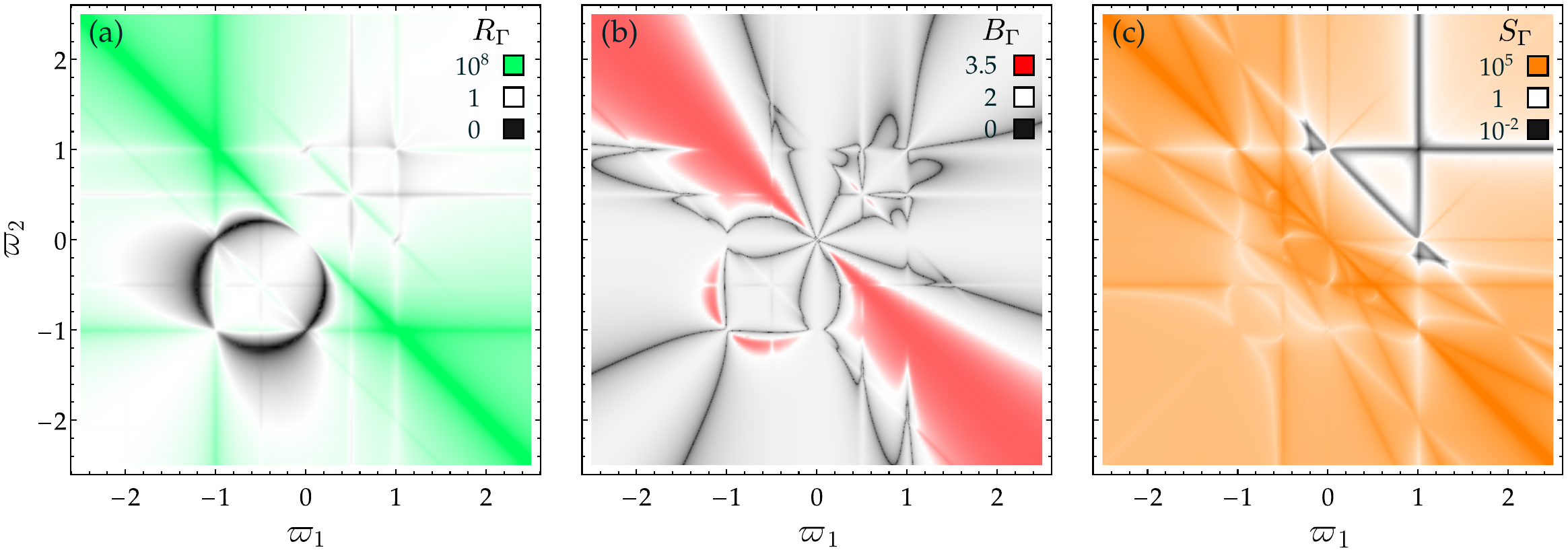}
  \caption{Violations of (a) Cauchy and (b) Bell inequalities in
    highly-detuned resonance fluorescence, as well as presence of (c)
    two-mode squeezing. Filtering harvests correlations. Sight of new
    physics beyond that discussed in the text is also
    apparent. Parameters are the same as for
    Fig.~\ref{fig:Sun7Jan183134CET2024}(f).}
  \label{fig:Sun7Jan184103CET2024}
\end{figure}

The above results invite one to seek new regimes of
strongly-correlated quantum emission, away from the spectral
peaks. There is much to prospect for, and various regions (photons of
different frequencies) provide us with different types of light. To
highlight this point, we compute various quantifiers of nonclassical
correlations, which can be applied to frequency-filtered
photons~\cite{sanchezmunoz14b}. We consider here the case of
highly-detuned resonance fluorescence, since it was not previously
considered in this context, and for being both a clean and fruitful
source of correlations. We also add a new quantifier for two-mode
squeezing, given the emphasis we have given to its role, also
clarifying that while we have compellingly established the importance
of squeezing in the broad phenomenology, we are far from having
exhausted the topic.  Figure~\ref{fig:Sun7Jan184103CET2024}(a) shows
the violation of Cauchy-Schwarz inequalities when~$R>1$ where
\begin{equation}
  \label{eq:Sun7Jan184613CET2024}
  R \equiv \big[ \g{2}_{12} \big]^2 / \big[\g{2}_{11} \g{2}_{22}\big] \,,
\end{equation}
while panel~(b) shows the violation of Bell's inequalities 
when~$B>2$ where, this time:
\begin{equation}
  \label{eq:Sun7Jan184711CET2024}
  B \equiv \sqrt{2} \bigg| \frac{\av{\varsigma_1^{\dagger 2} \varsigma^2_1}+\av{\varsigma_2^{\dagger 2} \varsigma^2_2}-4 \av{\varsigma_1^{\dagger}\varsigma_2^{\dagger} \varsigma_2 \varsigma_1}-\av{\varsigma_1^{\dagger 2} \varsigma^2_2}-\av{\varsigma_2^{\dagger 2} \varsigma^2_1}}{\av{\varsigma_1^{\dagger 2} \varsigma^2_1}+\av{\varsigma_2^{\dagger 2} \varsigma^2_2} + 2 \av{\varsigma_1^{\dagger}\varsigma_2^{\dagger} \varsigma_2 \varsigma_1}} \bigg| \,.
\end{equation}
Panel~(c) shows two-mode squeezing, which is present if and only if
$S > 1$, where~\cite{hillery89a}:
\begin{equation}
  \label{eq:Sun7Jan184818CET2024}
  S \equiv \frac{|\av{\varsigma_1^2 \varsigma_2^2}-\av{\varsigma_1 \varsigma_2}^2|}{  \av{\varsigma_1^{\dagger}\varsigma_2^{\dagger} \varsigma_2 \varsigma_1} 
- |\av{\varsigma_1} \av{\varsigma_2}|^2} \,.
\end{equation}
All these quantities are plotted in
Fig.~\ref{fig:Sun7Jan184103CET2024}. From the Cauchy-Schwarz
inequality, one can see again the region of no-coincidence emission,
where the system stops emitting at the two-photon level, which is the
black circle. This quantity might be even more suitable to identify
the region of no two-photon emission, being a more properly normalized
version of two-photon correlations. It generalizes to two-photon
emission the spectral line elimination via quantum interferences in
spontaneous emission~\cite{zhu96a}. The green regions show
nonclassical emission. Since the main diagonal is white, it means that
the emission at any frequency from the spectrum is classical, which is
also the case at resonance~\cite{sanchezmunoz14b}. Both the main
leapfrog and photons involving the virtual peak are strongly quantum
correlated, and are likely valuable resources for quantum emitters of
a new type. The Bell inequality is, interestingly, of a quite
different character, being mainly attached to the main leapfrog (a bit
as well to the circle of no-emission, surprisingly) where, unlike
other quantifiers that behave more like resonances, it spreads and
increases as one gets away from the triplet. Those are features that
we cannot currently explain. Finally, Panel~(c) which indicates
two-mode squeezing is probably the most interesting as it first
confirms that this is a feature to be found almost everywhere in
detuned resonance fluorescence, except in the vicinity of the real
peaks. More interestingly, however, new lines whose slope betray other
photon combinations ($2\varpi_1+\varpi_2=k$ and~$\varpi_1+2\varpi_2=k$
where~$k=-1$ or~$0$) reveal that there is rich additional physics that
we have not yet touched upon, possibly involving squeezed correlations
of higher photon numbers. Even at the qualitative level of explaining
the basic but strong features that shape the landscape, we find
evidence of more complicated processes inherited from beyond
two-photon physics. Some of them have been discussed through $\g{n}$
correlators at resonance~\cite{lopezcarreno17a}.  The features not yet
accounted for from the quantifiers discussed in this Section are
reported in Fig.~\ref{fig:Sun11Feb202132CET2024}(c) as thin lines,
requiring new physics.  In its pursuit, we could indeed, using the
same approach and techniques, but computing other observables,
characterize other types of correlations, such as three-mode
squeezing, other types of entanglement, quadrature-based tripartite
inseparability~\cite{armstrong15a,shalm13a}, etc. It is clear that
there remains much to explore, understand and turn into devices, even
with the simplest problem of quantum optics, while the same approach
can be applied to more complicated systems, from cosmic radiation to
atoms and molecules passing by condensed matter and other types of
quantum emitters.

\section{Conclusion and Outlook}

We have provided a detailed, and essentially analytical, picture of
the two-photon physics of resonance fluorescence. Maybe the most
important message is that this should be contemplated on its own,
independently from one-photon observables, however ingrained is one's
attachment to intensity, signal and photoluminescence spectra. This
text was written at the invitation of the Royal Society to commemorate
the Newton International Fellowship awarded to one of us (EdV, in
2009). It brings together various of EdV's research lines which
started at this occasion with a theory of lasing that produced the
Mollow triplet with coherence provided by the emitter itself as
opposed to being brought from outside~\cite{delvalle10d}. This led her
to study in more details the coupling (and decoupling) of two-level
systems~\cite{delvalle10b} and how they develop and maintain
correlations in the steady state~\cite{delvalle11b}, sustaining a rich
span of different regimes interpolating between incoherent and
coherent~\cite{delvalle11a} from the interplay of quantum and
dissipative light-matter interactions.  The need to scrutinize as
completely as possible photon correlations from such platforms led to
the introduction of the frequency degree of freedom, this time as part
of a Humboldt Fellowship. This provides the first pillar on which
stands the edifice that explains the two-photon physics of resonance
fluorescence (and a wealth of other problems in its wake): the theory
of frequency-resolved multiphoton
correlations~\cite{delvalle12a,delvalle13a}. The other pillar is the
theory of quantum field admixtures~\cite{zubizarretacasalengua20a}.
Remarkably, while these have been pursued independently, they turn out
to be naturally and deeply interconnected, besides, on what is
possibly the simplest-possible nontrivial problem of quantum optics:
the coherently driven two-level system. We believe that this betrays
the extremely fundamental and universal character of these concepts.
The other Authors of this text, as we are sure would also our past
collaborators on these topics, concur that the surprisingly complex
and rich phenomenology that is being revealed, far from exhausting or
completing the description of the problem, is only making it even more
mysterious and unfathomable.  This evokes to us the aphorism of Newton
himself:

\begin{quotation}
  «I seem to have been only like a boy playing on the sea-shore, and
  diverting myself in now and then finding a smoother pebble or a
  prettier shell than ordinary, whilst the great ocean of truth lay
  all undiscovered before me.»
\end{quotation}

While we have for brevity, simplicity and current experimental
interest, focused on two-photon physics, the above approach is
completely general and can, indeed inevitably will, be generalized to
multiphoton correlations and $n$-mode squeezing.  Spheres of
antibunching~\cite{lopezcarreno17a} have already been spotted in
three-photon correlation spectra and it is obvious that resonance
fluorescence abounds with three-mode squeezing that remain to be
characterized, measured and exploited. Among other compelling
immediate continuations of this work, we can mention i) seizing
additional control and tuneability of the correlations by externally
adjusting the interferences with a laser~\cite{lopezcarreno18a}, ii)
using strongly-correlated two-photon spectral locations for
quantum-spectroscopic applications~\cite{mukamel20a} or iii)
Purcell-enhancing them to realize new types of devices with
high-purity and strong signal~\cite{sanchezmunoz14a}. There are
definitely still other and possibly even more exciting prospects.  At
any rate, it is clear that, although inconspicuous at the one-photon
level, there is a rich and varied two-photon physics, which takes
place everywhere.

\ack{We acknowledge discussions with and interest from colleagues
  throughout the years, in particular Juan Camilo L\'opez Carre{\~n}o,
  Carlos S\'anchez Montenegro, Alejandro Gonz\'alez Tudela, Carlos Ant\'on
  Solanas, Carlos Tejedor, Sang Kyu Kim and Kai M\"uller. EdV
  acknowledges support from the CAM Pricit Plan (Ayudas de Excelencia
  del Profesorado Universitario), TUM-IAS Hans Fischer Fellowship and
  projects AEI/10.13039/501100011033 (2DEnLight) and Sinérgico CAM
  2020 Y2020/TCS-6545 (NanoQuCo-CM). FPL acknowledges the HORIZON
  EIC-2022-PATHFINDERCHALLENGES-01 HEISINGBERG project 101114978. This
  work was written at the invitation of EdV's Newton Fellowship, which
  is also most gratefully acknowledged.}


\bibliographystyle{naturemag}
\bibliography{sci,Books,arXiv}

\end{document}